\documentclass[12pt,preprint]{aastex}

\newcommand{\ld}{\lambda/D}
\newcommand{\rhonaught}{\rho_{\mbox{\scriptsize iwd}}}
\newcommand{\rhoone}{\rho_{\mbox{\scriptsize owd}}}
\newcommand{\hoverra}{\alpha}    
\newcommand{\hoverrb}{\alpha}  
\newcommand{\hovertwor}{\frac{\alpha}{2}}  

\usepackage[intlimits]{amsmath}

\shorttitle{Spiderweb Masks}
\shortauthors{Vanderbei et al.}

\begin{document}


\title{Spiderweb Masks for High-Contrast Imaging}


\author{Robert J. Vanderbei}
\affil{Operations Research and Financial Engineering, 
Princeton University}
\email{rvdb@princeton.edu}

\author{David N. Spergel}
\affil{Astrophysical Sciences, Princeton University}
\email{dns@astro.princeton.edu}

\and

\author{N. Jeremy Kasdin}
\affil{Mechanical and Aerospace Engineering, Princeton University}
\email{jkasdin@princeton.edu}


\begin{abstract}
Motivated by the desire to image exosolar planets,
recent work by us and others has shown that high-contrast imaging
can be achieved using specially shaped pupil masks.  To date,
the masks we have designed have been symmetric with respect to a 
cartesian coordinate
system but were not rotationally invariant, thus requiring that
one take multiple images at different angles of rotation about the
central point in order to obtain high-contrast in all directions.
In this paper, we present a new class of masks that have rotational
symmetry and provide high-contrast in all directions with just one
image.  These masks provide the required $10^{-10}$ level of contrast 
to within $4 \lambda/D$, and in some cases $3 \ld$,
of the central point, which is deemed necessary
for exo-solar planet finding/imaging.  They are also well-suited
for use on ground-based telescopes, and perhaps NGST too, since
they can accommodate central obstructions and associated support spiders.
\end{abstract}




\section{Introduction}
\label{sec:intro}

With more than 100 extrasolar Jupiter-sized
planets discovered in just the last decade, 
there is now great interest in discovering
and characterizing Earthlike planets.  To this end, NASA is planning to
launch a space-based telescope, 
called the {\em Terrestrial Planet Finder (TPF)},
sometime in the middle of the next decade.  This telescope, which
will ultimately be either an interferometer or a coronagraph, will be
specifically designed for high-contrast imaging.  Earlier studies 
(\citet{ref:Brown}) 
indicate that a $4$m class coronagraph ought to be able to discover
about 50 extrasolar Earth-like planets if the telescope can provide contrast
of $10^{-10}$ at a separation of $3\ld$ and that a $4 \times 10$m class
telescope ought to be able to discover about 150 such
planets if it can provide the same contrast at a separation of $4\ld$.

One of the most promising design concepts for high-contrast imaging
is to use pupil masks for diffraction
control.
So far, work in this direction 
(see \citet{ref:kasdin,ref:Brown,ref:Spergel,KVSL02b,KVSL02}) 
has focused on optimizing masks that are not rotationally symmetric
and thus provide the desired contrast only in a narrow annular sector around a
star, but at fairly high throughput.  
A full investigation of a given star then requires multiple images
where the mask is rotated between the images so as to ultimately image
all around the star.  In this paper, we propose a new class of rotationally
symmetric pupil masks, which are attractive because they do not require
rotation to image around a star.  Such masks consist of concentric rings.  As we
show, they can provide the desired contrast with a reasonable amount of
light throughput.  However, it is not clear {\em a priori}\ how to support
a concentric-ring mask.  While the simplest mechanism would be 
to lay the rings on a glass plate, even a tiny amount of scatter
from the glass could destroy the high contrast.  It is better to use
an opaque structure for support, maintaining the purely binary nature of the
mask.  We show here that a certain type of
spider support can be used without adversely affecting the contrast in the
desired working range.

The paper is organized as follows.  In the next section, we show how to design
high-contrast concentric-ring masks as a two-step optimization process.
The first step is to solve a linear programming problem for an optimal
apodization.  The second step exploits the ``bang--bang'' nature of the
optimal apodization to create a starting point for a local nonlinear search
for an optimal mask.  With masks in hand, Section \ref{sec:spiders} 
is devoted to
considering the impact of support spiders on the point spread function and in
particular on finding spiders that preserve the high-contrast region.
Finally, we discuss how to design masks with central obstructions.
Such masks, with their spiders, could be useful for telescopes with 
on-axis designs
such as NGST and a variety of ground-based telescopes.  For these,
the mask's spiders could be used to block the spiders supporting the secondary.

\section{Concentric-Ring Masks}
\label{sec:masks}

The image-plane electric field produced by 
an on-axis point source and an apodized aperture defined
by a circularly-symmetric apodization function $A(\sqrt{x^2 + y^2})$ is given
by 
\begin{equation}
    E(\xi,\zeta) = 
    \iint_S e^{-2 \pi i (x \xi + y \zeta )} A\left(\sqrt{x^2+y^2}\right) dx dy, 
\end{equation}
where
\begin{equation}
    S = \displaystyle \left\{ (x,y): \;
    		          0 \le r(x,y) \le 1/2, \;
                          \theta(x,y) \in \left[0, 2\pi \right]
		       \right\},
\end{equation}
and $r(x,y)$ and $\theta(x,y)$ denote the polar coordinates associated with
point $(x,y)$.  
Here, and throughout the paper, $x$ and $y$ denote coordinates in the pupil
plane measured in units of the aperature $D$ and $\xi$ and $\zeta$ denote
angular (radian) deviation from on-axis measured in units of wavelength
over aperture ($\lambda/D$) or, equivalently, physical distance in the image
plane measured in units of focal-length times wavelength over aperture 
($f \lambda/D$).
If the apodization function $A()$ takes only the values $0$ 
and $1$, then the apodization can be realized as an aperture mask.

For circularly-symmetric apodizations and masks, 
it is convenient to work in polar
coordinates.  To this end, let $r$ and $\theta$ denote polar coordinates in
the pupil plane and let $\rho$ and $\phi$ denote the image plane coordinates:
\begin{equation}
    \begin{array}{rclrcl}
        x & = & r \cos \theta \qquad \qquad   \xi & = & \rho \cos \phi \\
        y & = & r \sin \theta \qquad \qquad \zeta & = & \rho \sin \phi .
    \end{array}
\end{equation}
Hence,
\begin{eqnarray}
 x \xi + y \zeta &=& r \rho (\cos \theta \cos \phi + \sin \theta \sin \phi)\\
                 &=& r \rho \cos(\theta-\phi)  .
\end{eqnarray}
The electric field in polar coordinates depends only on $\rho$ and is given by
\begin{eqnarray}
    E(\rho) 
    &=&  \label{eq1}
    \int_0^{1/2} \int_0^{2\pi} 
        e^{-2 \pi i r \rho \cos(\theta-\phi)} A(r) 
    r d\theta dr, \\
    &=&  \label{eq2}
    2 \pi \int_0^{1/2} J_0(2 \pi r \rho) A(r) r dr,
\end{eqnarray}
where $J_0$ denotes the $0$-th order Bessel function of the first kind.
Note that the mapping from apodization function $A$ to electric field $E$
is linear.  Furthermore, the electric field in the image plane is real-valued
(because of symmetry) and its value at $\rho = 0$
is the {\em throughput} of the apodization:
\begin{equation}
    E(0) = 2 \pi \int_0^{1/2} A(r) r dr.
\end{equation}

The {\em point spread function} (psf) is the square of the electric field.
The contrast requirement is that the psf in the dark region be $10^{-10}$ of
what it is at its center.  Because the electric field is real-valued, it
is convenient to express the contrast requirement in terms of it rather than
the psf, resulting in a field requirement of $\pm 10^{-5}$.

The apodized pupil that maximizes throughput subject to contrast constraints
can be formulated as an infinite dimensional linear programming problem:
\begin{equation}
    \begin{array}{ll}
        \mbox{maximize } & E(0) \\
	\mbox{subject to } &
	    \setlength{\arraycolsep}{0.1em}
	    \begin{array}[t]{rll}
	    \displaystyle
	        -10^{-5} E(0) & \le E(\rho) \le 10^{-5} E(0), &
	            \qquad \rhonaught \le \rho \le \rhoone , \\
		0 & \le A(r) \le 1, &
		    \qquad 0 \le r \le 1/2 ,
	    \end{array}
    \end{array}
\end{equation}
where $\rhonaught$ denotes a fixed {\em inner working distance} and $\rhoone$ a
fixed {\em outer working distance}.
Discretizing the sets of $r$'s and $\rho$'s and replacing the integrals with
their Riemann sums, the problem is approximated by a finite dimensional linear
programming problem that can be solved to a high level of precision
(see, e.g., \citet{Van01}).

\begin{figure}
\begin{center} 
\includegraphics[width=3.2in]{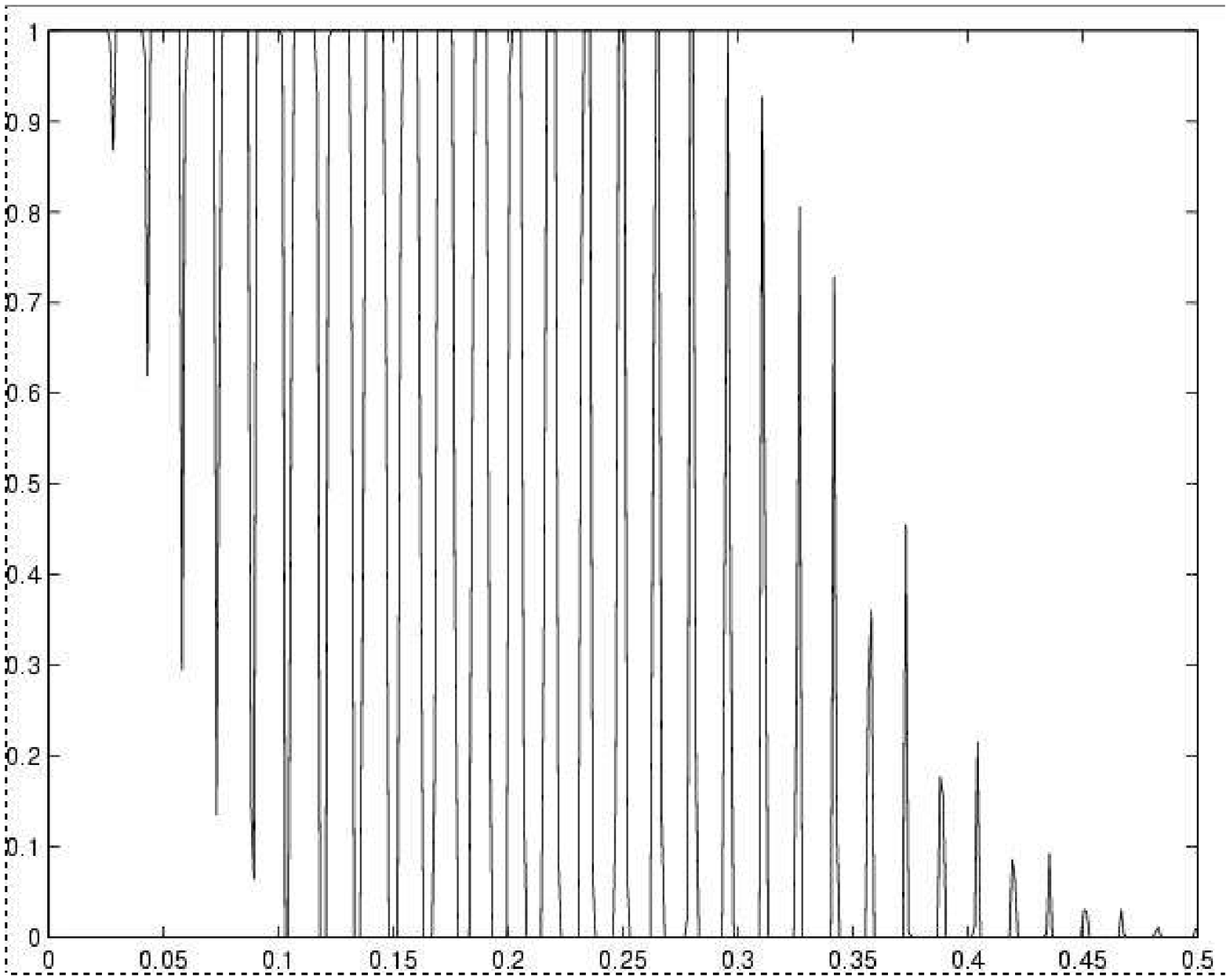} 
\includegraphics[width=3.2in]{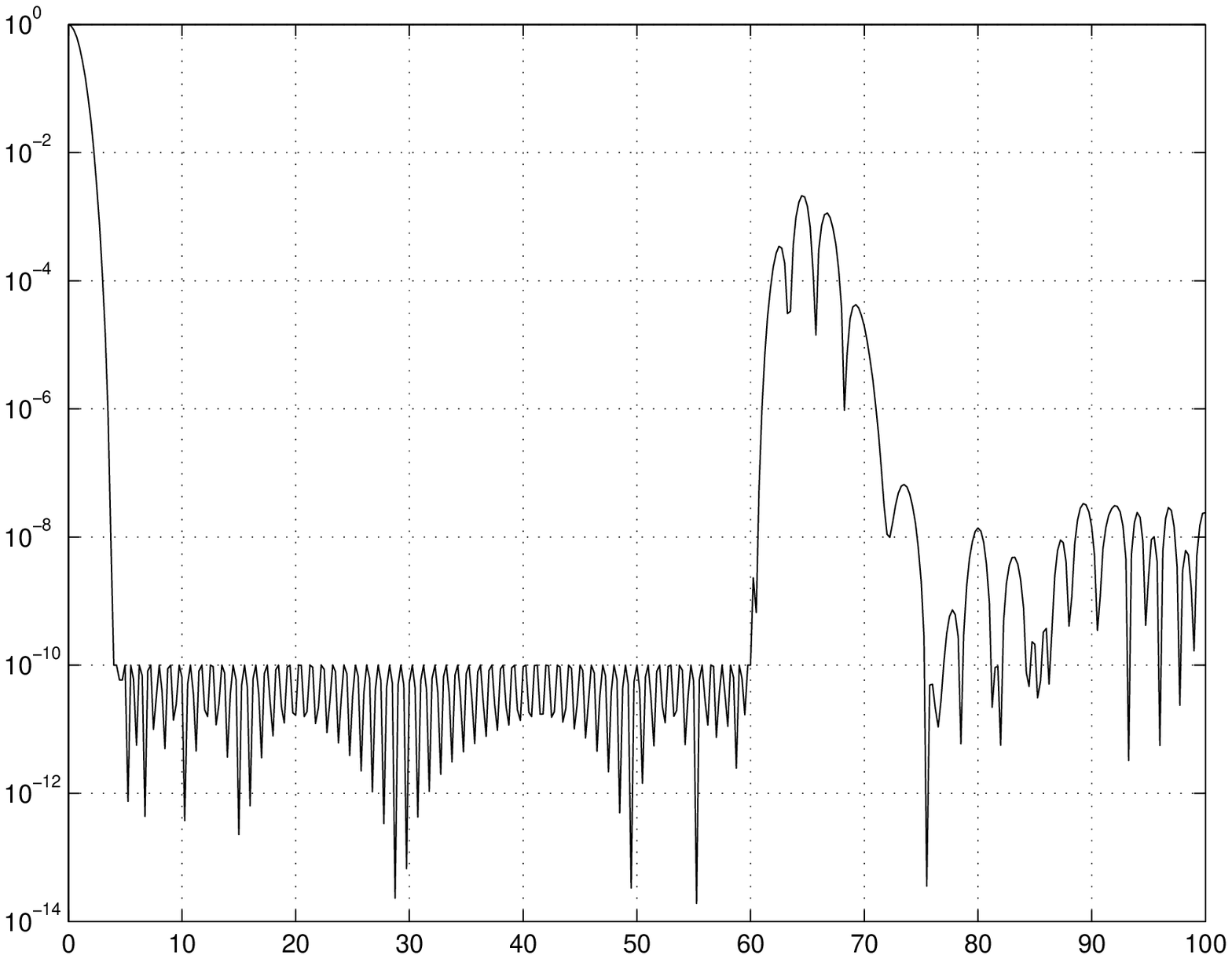} 
\end{center}
\caption{The optimal apodization for $\rhonaught = 4$ and $\rhoone = 60$
and the associated psf.
}
\label{fig:fig7}
\end{figure}

\begin{figure}
\begin{center} \includegraphics[width=2.5in]{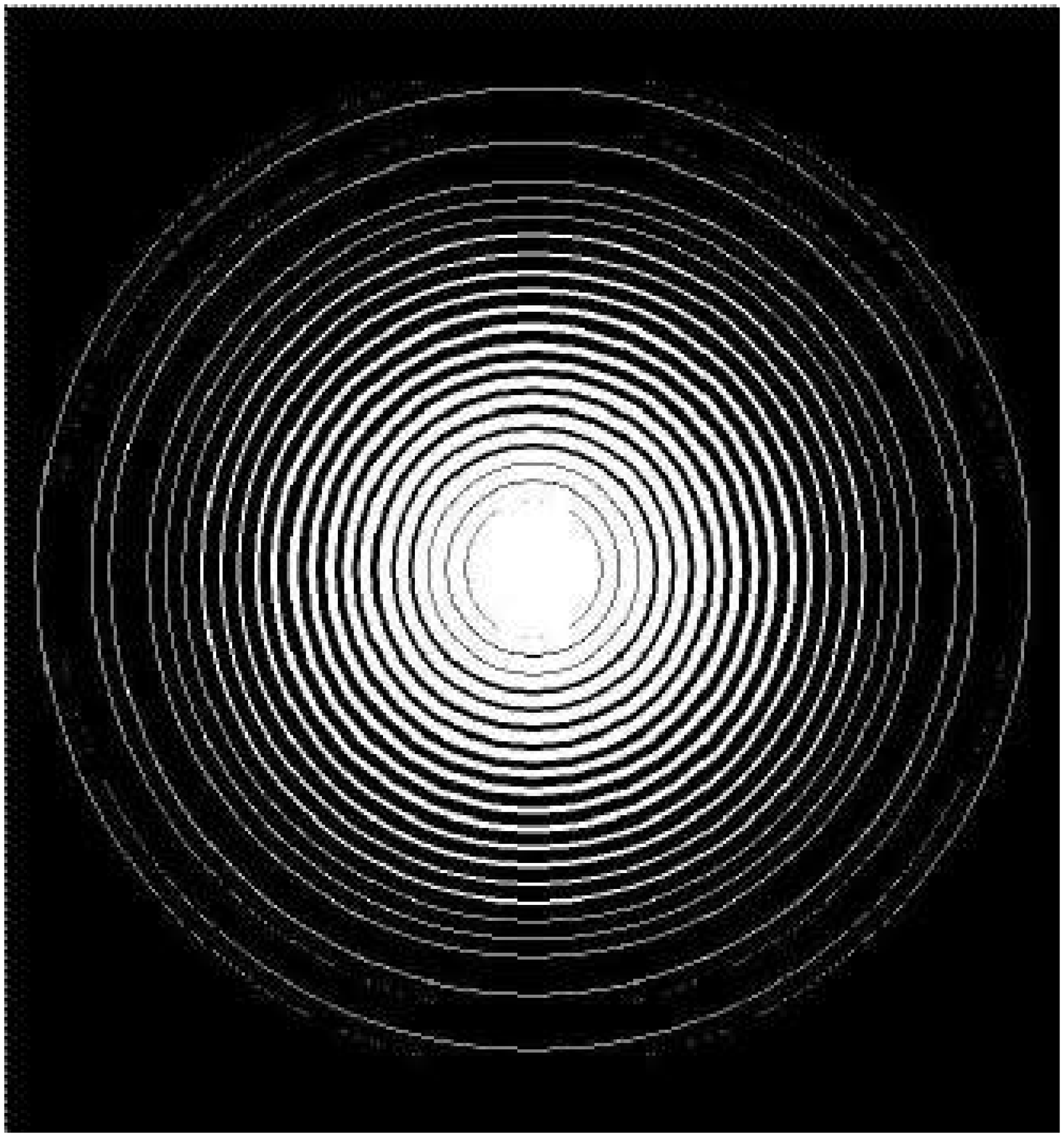} \end{center}
\begin{center} 
\hfill \includegraphics[width=2.0in]{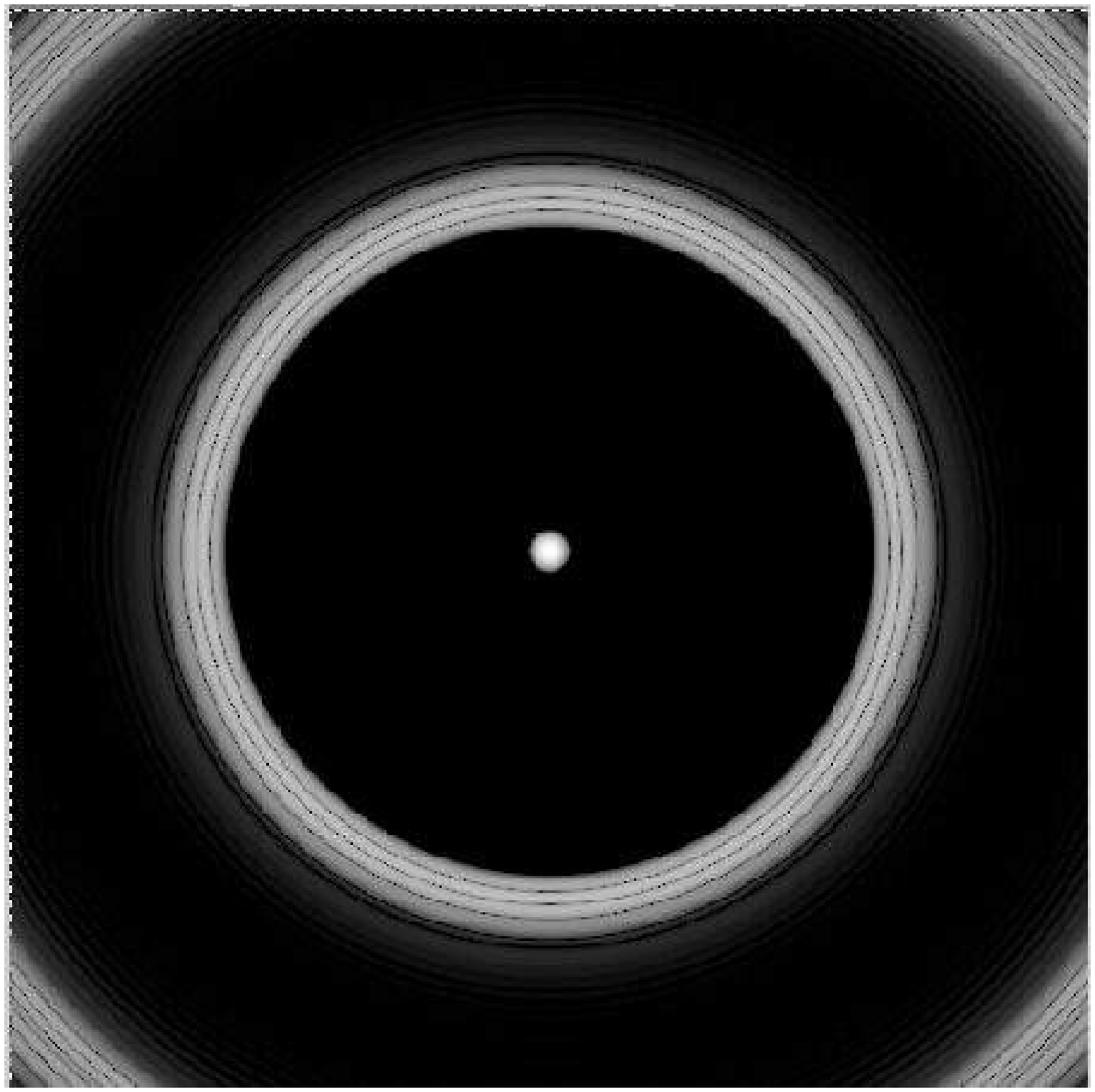}
\hfill \includegraphics[width=2.5in]{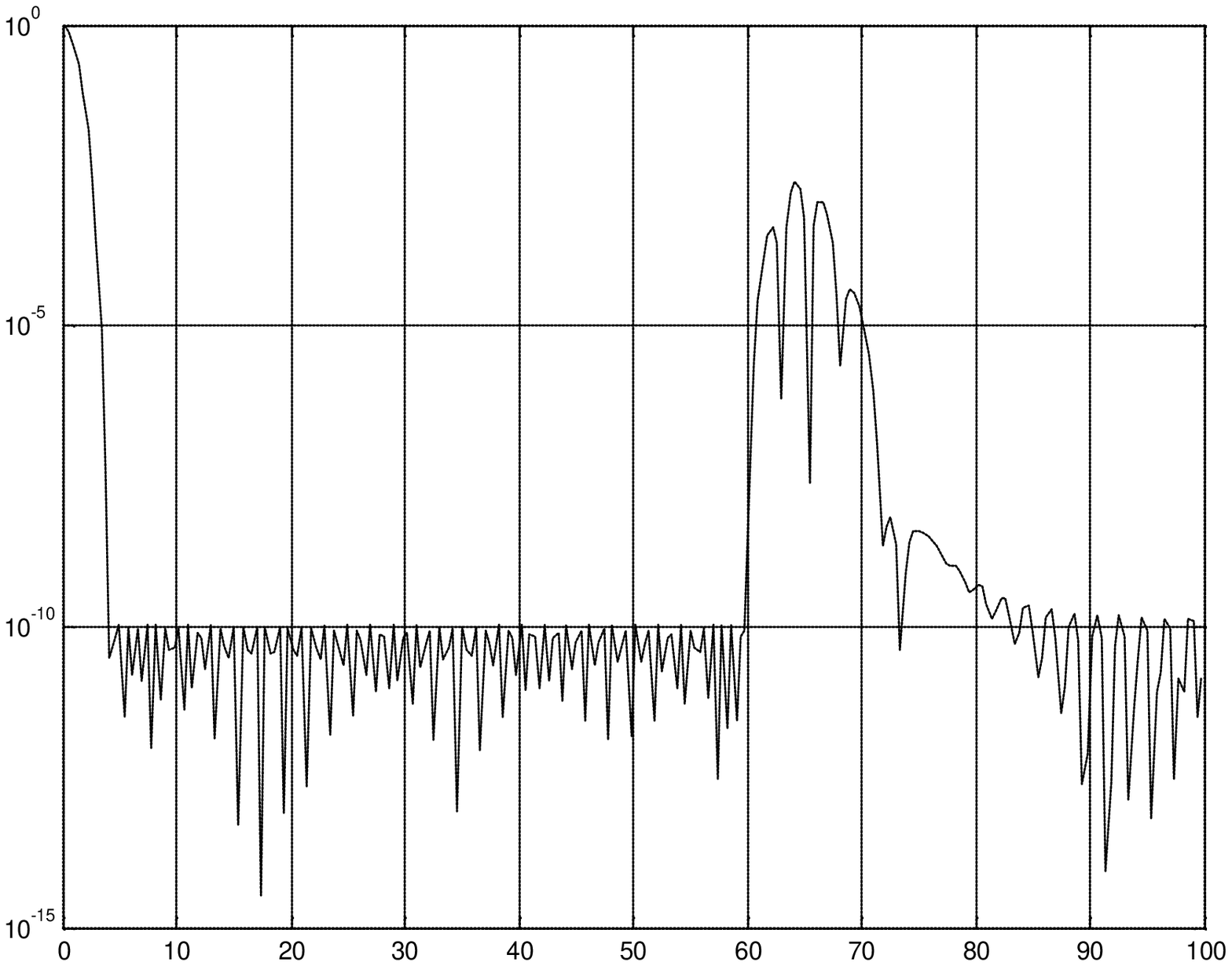}
\hfill ~ 
\end{center}
\caption{{\em Top.} A concentric-ring mask designed to provide high-contrast,
$10^{-10}$, from $\ld = 4$ to $\ld=60$.  Throughput is $0.134 = 17.00\%$.
{\em Bottom.} The associated psf.
}
\label{fig:fig1}
\end{figure}

The solution obtained for $\rhonaught = 4$ and $\rhoone = 60$ is shown in Figure
\ref{fig:fig7}.  Note that the solution is of a bang-bang type.  That is,
the apodization function is mostly $0$ or $1$ valued.  This suggests looking
for a mask that is about as good as this apodization.
Such a mask can be found by solving the following nonlinear optimization 
problem.  A mask consists of a set of
concentric opaque rings, formulated in terms of
the inner and outer radii of the openings between the rings:
\begin{eqnarray}
    \; [r_0,r_1] && \mbox{first opening} \\
    \; [r_2,r_3] && \mbox{second opening} \\
    \; [r_4,r_5] && \mbox{third opening} \\
           & \vdots & \\
    \; [r_{2m-2},r_{2m-1}] && \mbox{$m$-th opening} 
\end{eqnarray}
With this notation, the formula for $E(\rho)$ given in \eqref{eq2} 
can be rewritten as a sum of integrals over these openings:
\begin{eqnarray}
    E(\rho) 
    &=& 
    2 \pi \sum_{k=0}^{m-1} \int_{r_{2k}}^{r_{2k+1}} J_0(2 \pi r \rho) r dr, \\
    &=& 
    \frac{1}{\rho}
    \sum_{k=0}^{m-1} \left( 
                        r_{2k+1}J_1(2\pi r_{2k+1}\rho) 
			-
                        r_{2k}J_1(2\pi r_{2k}\rho) 
                     \right)
\end{eqnarray}
Treating the $r_k$'s as variables and using this new expression for the
electric field, the mask design problem becomes:
\begin{equation}
    \begin{array}{ll}
        \mbox{maximize } & 
		\displaystyle
	        \pi \sum_{k=0}^{m-1} \left( r_{2k+1}^2 - r_{2k}^2 \right) \\
	\mbox{subject to } &
	        -10^{-5} E(0) \le E(\rho) \le 10^{-5} E(0), 
	            \qquad \rhonaught \le \rho \le \rhoone , \\
		& 0 \le r_0 \le r_1 \le \cdots \le r_{2m-1} \le 1/2 .
    \end{array}
\end{equation}
This problem is a nonconvex nonlinear optimization problem and hence
the best hope for solving it in a reasonable amount of cpu time is to
use a ``local-search'' method starting the search from a solution that
is close to optimal.  The bang-bang solution from the 
linear programming problem can be used to generate a starting solution.
Indeed, the discrete solution to the linear programming problem can
be used to find the inflection points of $A$ which can be used as initial
guesses for the $r_k$'s.  
The particular local-search method we used is the first-author's {\sc loqo}
optimizer, which is described in \citet{SOR9708}.
Figures \ref{fig:fig1} and \ref{fig:fig3}
show optimal concentric-ring masks computed using
an inner working distance of $4$
and two different choices of outer working distances ($40$ and $60$).  

There is great interest in decreasing the inner working distance as this
rapidly increases the number of possible planets one could discover via TPF.
Or, put another way, it allows one to study the same set of targets but with a
smaller aperture instrument.
Using $\rhonaught = 3.5$, we first tried to solve the linear programming
problem with $\rhoone = 10$.  There is no solution.  We next tried 
$\rhoone = 7$ and found a solution to both the linear and nonlinear
optimization problems.  The resulting mask is shown in Figure
\ref{fig:fig5}.  

Figure \ref{fig:fig10} shows a mask with $\rhonaught = 3$
and $\rhoone = 4.8$.  The linear program with $\rhoone = 5.0$ is infeasible.
We have even created a mask with $\rhonaught=2.5$ and $\rhoone = 3.1$.  In the
interest of space, we don't show this last mask.

Of course, the obvious issue is ring support.
How can the rings be supported without compromising the high-contrast area of
the psf, without significantly reducing 
throughput, and without using glass which would introduce unacceptable
scatter?
The simplest answer is to use a ``spider'' similar to those used, for example,
to support a secondary mirror.  However, spiders create
diffraction spikes which can destroy the high-contrast.  In the next section,
we investigate the effect of spiders and show how to make them in such a way
as to preserve high-contrast.

\section{Using Spiders for Support}
\label{sec:spiders}

In this section, we study the impact that ring-connecting
spiders have on the masks computed in the previous section.  
Normally, one thinks of a spider as a uniform-width support piece used to
hang things.  However, there is no specific need that these supports be of
uniform width.  In fact, since the masks will be made by electroforming
technology, such a restriction to uniform width is completely unnecessary.
Instead of constant-width spiders, we study constant-angular-width spiders.
That is, each spider vane is a narrow sector, starting from a point at the
center and flaring out with radius.
Incorporating $N$ such spiders uniformly spaced, 
the electric field, expressed in polar coordinates, is given by
\begin{equation} \label{eq3}
    E(\rho,\phi) 
    =
    \displaystyle
    \iint_S e^{-2 \pi i r\rho \cos(\theta-\phi)} A(r) r dr d\theta
\end{equation}
where
\begin{equation}
    A(r) = \sum_{k=0}^{m-1} 1_{[r_{2k},r_{2k+1}]} (r),
\end{equation}
\begin{equation}
    S = \displaystyle \left\{ (x,y): \;
    		          0 \le r(x,y) \le 1/2, \;
                          \theta(x,y) \in \Theta 
		       \right\},
\end{equation}
\begin{equation}
    \Theta = \bigcup_{n=0}^{N-1} 
		   \left[
			   \frac{2 \pi n}{N}     + \hovertwor,
			   \frac{2 \pi (n+1)}{N} - \hovertwor
		   \right] ,
\end{equation}
$\hoverra$ denotes the width of a vane in radians, the notation
$[a,b]$ denotes the interval on the real line from $a$ to $b$,
and the notation $1_{[a,b]}$ denotes the function that is one
on the interval $[a,b]$ and zero elsewhere.

The integral in equation \eqref{eq3} can be expressed in terms
of Bessel functions using the Jacobi-Anger expansion (see, e.g., 
\citet{AW00} p. 681):
\begin{equation}
    e^{i x \cos \theta} 
    =
    \sum_{m=-\infty}^{\infty} i^m J_m(x) e^{i m \theta} .
\end{equation}
Substituting into \eqref{eq3}, we get:
\begin{eqnarray}
    E(\rho,\phi)
    &=&
    \displaystyle
    \iint_S 
       \sum_m i^m J_m(-2 \pi r \rho) e^{im(\theta-\phi)} A(r) 
    r dr d\theta \\
    &=&
    \displaystyle
    \int_0^{1/2} 
       \sum_m i^m J_m(-2 \pi r \rho) e^{-im\phi} 
       \left( \int_{\Theta} e^{i m \theta} d\theta \right)
       A(r) 
    r dr 
\end{eqnarray}
The integral over $\Theta$ is easy to compute:
\begin{eqnarray}
    \int_{\Theta} e^{i m \theta} d \theta
    &=&
    \sum_{n=0}^{N-1} 
    \int_{\frac{2 \pi n}{N}+\hovertwor}^{\frac{2 \pi (n+1)}{N}-\hovertwor} 
        e^{im\theta} d\theta
    \\
    &=&
    \left\{
	\begin{array}{ll}
            2 \pi - N\hoverrb & \quad m = 0 \\
	                       - \frac{2}{j} \sin(jN\hovertwor)
			       	 & \quad m = jN, \; j \ne 0 \\
			       0 & \quad \mbox{otherwise} .
	\end{array}
    \right.
\end{eqnarray}
Substituting this result into \eqref{eq3}, yields
\begin{eqnarray}
    E(\rho,\phi)
    &=&
    \int_0^{1/2} J_0(-2\pi r \rho)(2\pi - N\hoverrb) A(r) r dr \\
    && - \sum_{j \ne 0}
    \int_0^{1/2} i^{jN} J_{jN}(-2\pi r \rho)e^{-ijN\phi} 
	                            \frac{2}{j} \sin(jN\hovertwor)
				    A(r) r dr .
\end{eqnarray}
Lastly, suppose that $N$ is even 
and use the fact that $J_{-m}(x) = J_m(-x) = (-1)^m J_m(x)$ to get the
following expansion for the electric field:
\begin{eqnarray}
    E(\rho,\phi)
    &=&
    (2\pi - N\hoverrb) \int_0^{1/2} J_0(2\pi r \rho) A(r) r dr \\
    && - 4 \sum_{j=1}^{\infty} 
        \int_0^{1/2} J_{jN}(2\pi r \rho)\cos(jN(\phi-\pi/2)) 
	                       \frac{1}{j} \sin(jN\hovertwor)
				A(r) rdr .
\end{eqnarray}

The first term, involving the integral of $J_0$,
is identical, up to a constant factor, to the formula for the electric
field in the absence of spiders.  Hence, the impact of the spiders is
two-fold: (a) the original electric field is reduced by $1 - N\hoverrb/(2\pi)$
which is just the fraction of the open area that remains uncovered by the
spiders,
and (b) there are additional terms containing higher order Bessel functions.

For large $N$, the effect of the higher-order Bessel terms becomes negligible
for small $\rho$.  Indeed, for $x \le \sqrt{4(m+1)}$, 
\begin{equation}
    0 \le J_m(x) \le \frac{(x/2)^{m+1}}{(m+1)!} ,
\end{equation}
(which itself follows easily from the alternating Taylor series expansion
of the $m$-th Bessel function: 
$J_m(x) = \sum_{l=0}^{\infty} (-1)^l (x/2)^{2l+m}/(l!(m+l)!)$).
From this we use the Schwarz inequality to estimate the magnitude of the
effect of the spiders:
\begin{equation}
\setlength{\arraycolsep}{0in}
\begin{array}{rcl}
    \displaystyle
    \left| \phantom{\int_0^{1/2}}4 \sum_{j=1}^{\infty} 
        \int_0^{1/2} J_{jN}(2\pi r \rho)\cos(jN(\phi \right.&-&\left. \pi/2)) 
			       \displaystyle
	                       \frac{1}{j} \sin(jN\hovertwor)
				A(r) rdr
				\phantom{\int_0^{1/2}}
				\right| \\
	&\le& \; \displaystyle
	      \frac{1}{2} \sum_{j=1}^{\infty} J_{jN}(\pi \rhoone) \\
	&\le& \; \displaystyle
	      \frac{1}{2} \sum_{j=1}^{\infty} 
	      \frac{1}{(jN+1)!} \left( \frac{\pi \rhoone}{2} \right)^{jN+1} ,
\end{array}
\end{equation}
for $\rhoone \le \sqrt{4(N+1)}/\pi$.
Since the last bound is dominated by $e^{\pi \rhoone/2}/2$, it follows from
the dominated convergence theorem that this last bound tends to zero as $N$
tends to infinity.

The convergence to zero of the terms in the sum on $j$ is very fast.
In fact, if $N$ is set large enough that 
$\max_{0 \le x \le \pi \rhoone} J_N(x) \le 10^{-5}$, 
then the $j=1$ term dominates the sum of all the higher-order terms and
is itself dominated by the $J_0$ term.  Figure \ref{fig:fig8} shows a plot of
$J_{50}$, $J_{100}$, and $J_{150}$.  These three Bessel functions first reach
$10^{-5}$ at $35.2$, $81.0$, and $128.1$, respectively.

\begin{figure}
\begin{center} 
\hfill \includegraphics[width=2.0in]{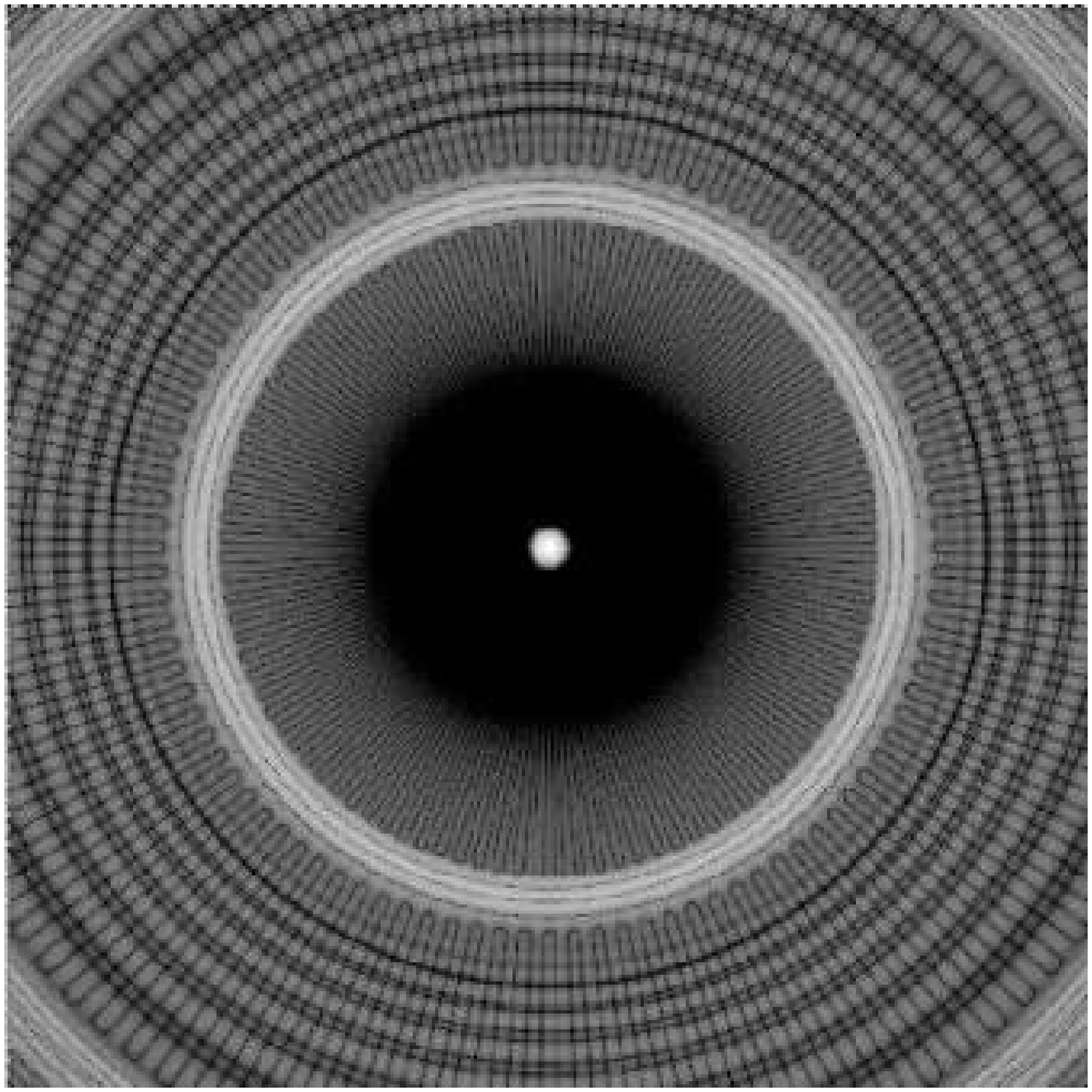}
\hfill \includegraphics[width=2.5in]{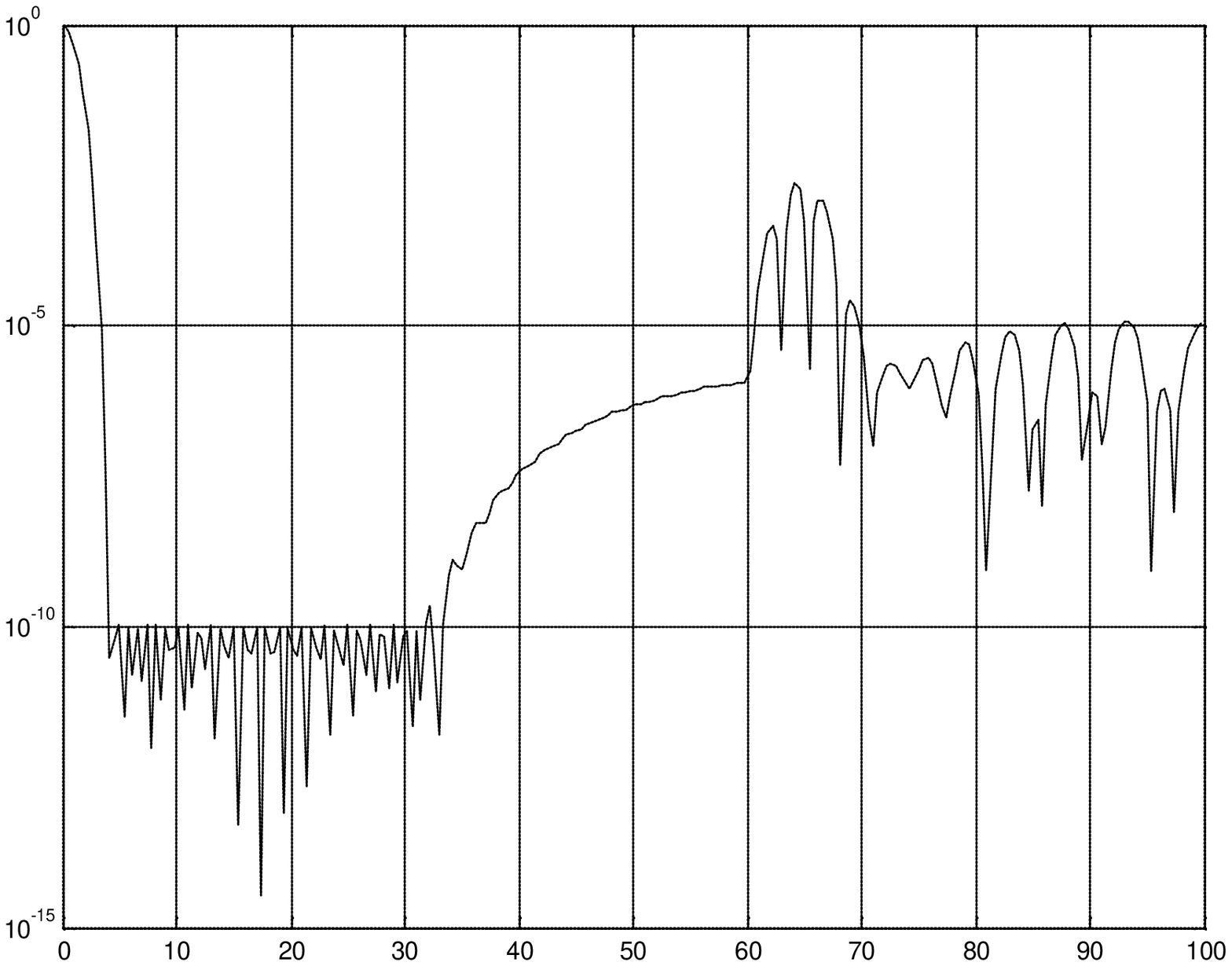}
\hfill ~ \\
\hfill \includegraphics[width=2.0in]{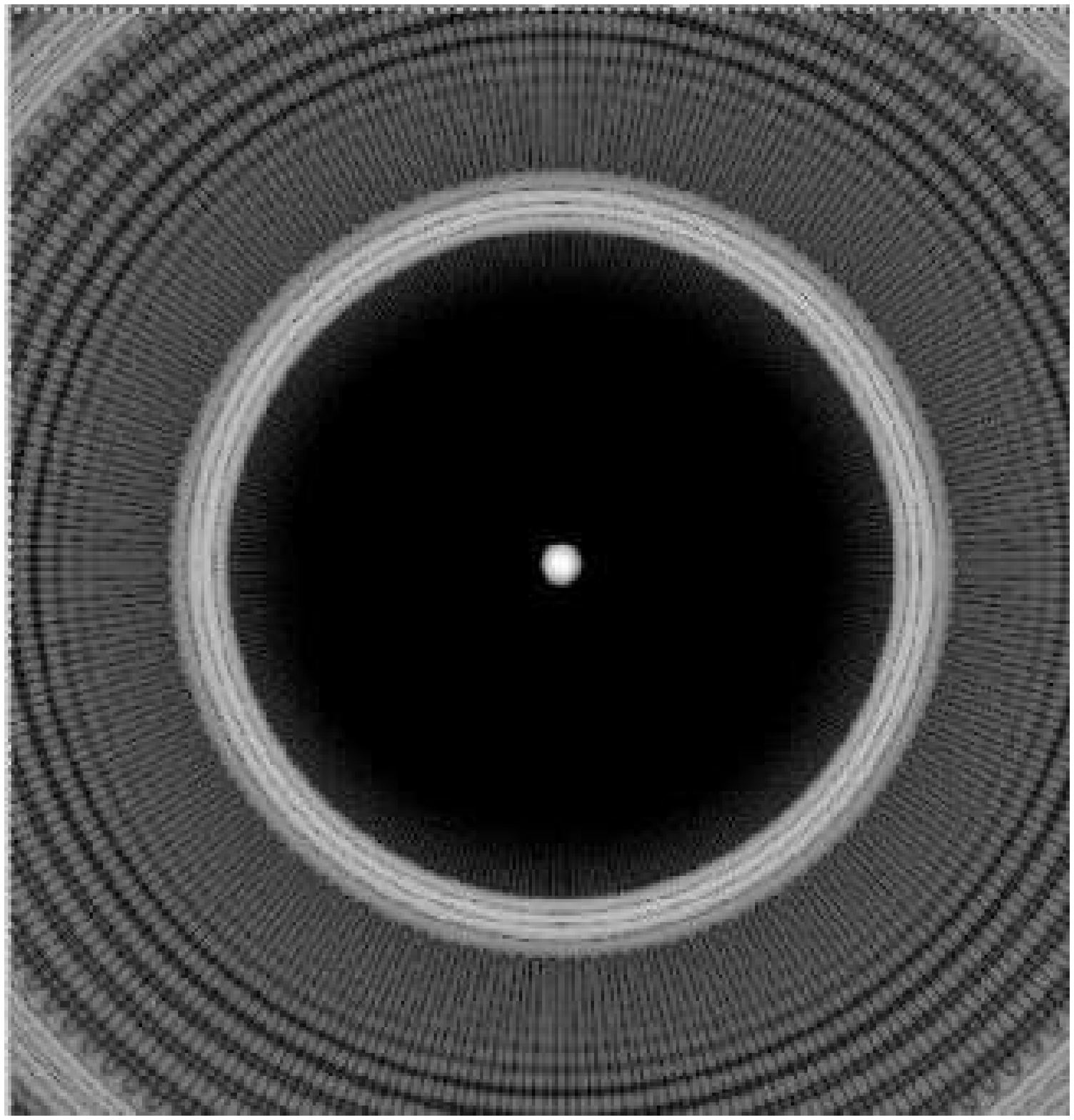}
\hfill \includegraphics[width=2.5in]{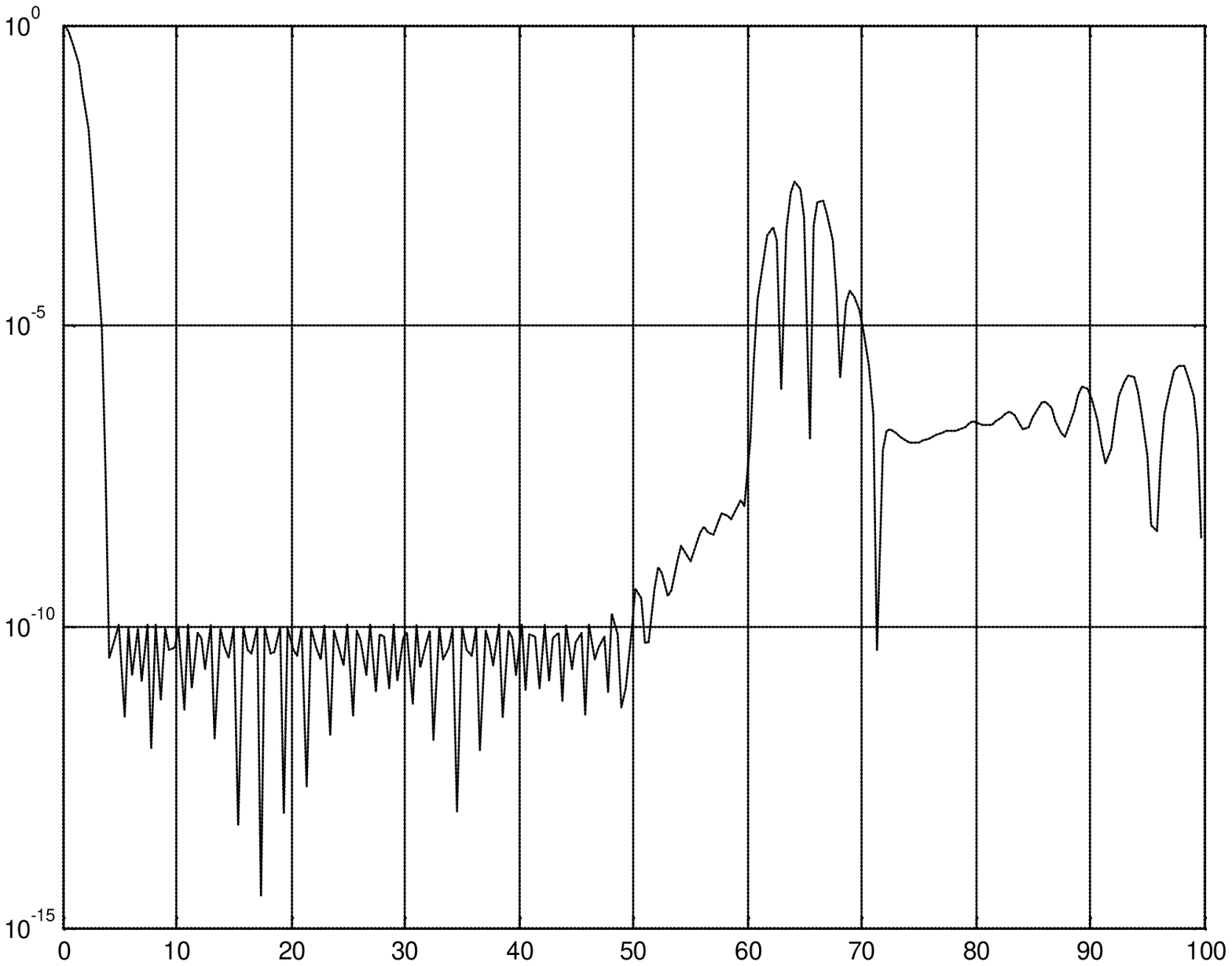}
\hfill ~ \\
\hfill \includegraphics[width=2.0in]{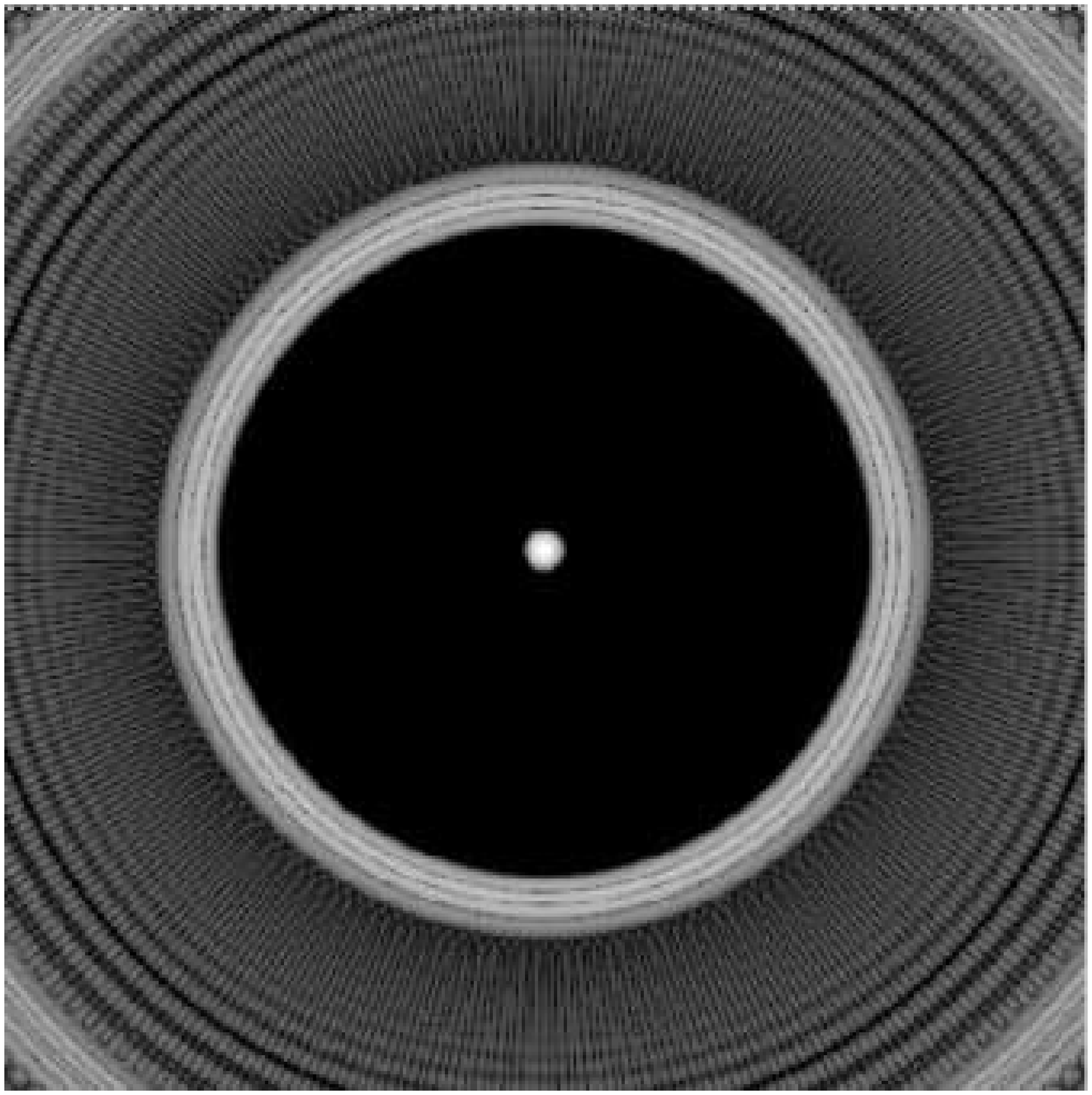}
\hfill \includegraphics[width=2.5in]{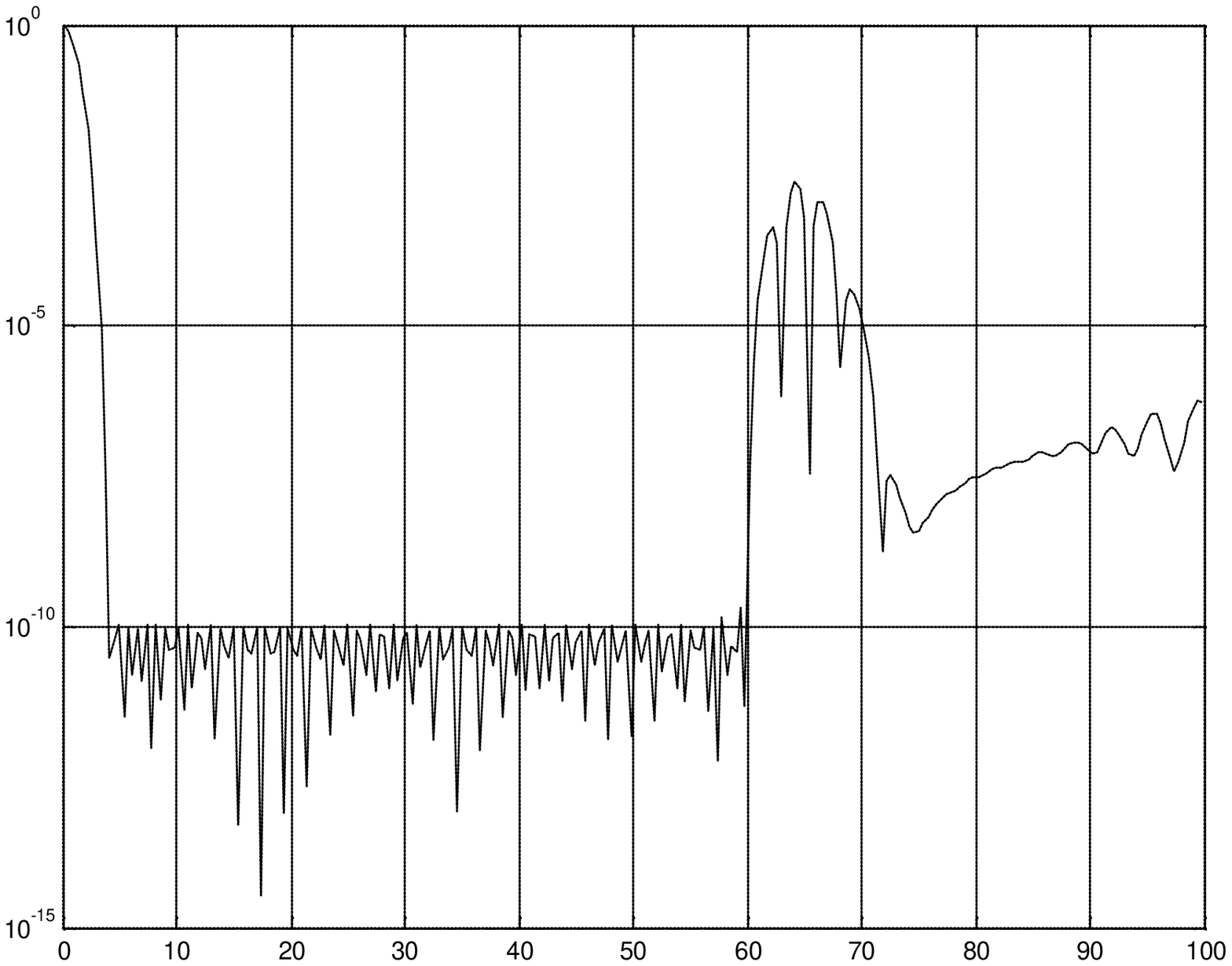}
\hfill ~ 
\end{center}
\caption{Psf for mask in Figure \ref{fig:fig1} with spiders.  
{\em Top Row.} $100$ spiders each spanning $0.005$ radians.
{\em Second Row.} $150$ spiders each spanning $0.003$ radians.
{\em Third Row.} $180$ spiders each spanning $0.0025$ radians.
}
\label{fig:fig2}
\end{figure}

\begin{figure}
\begin{center} \includegraphics[width=2.5in]{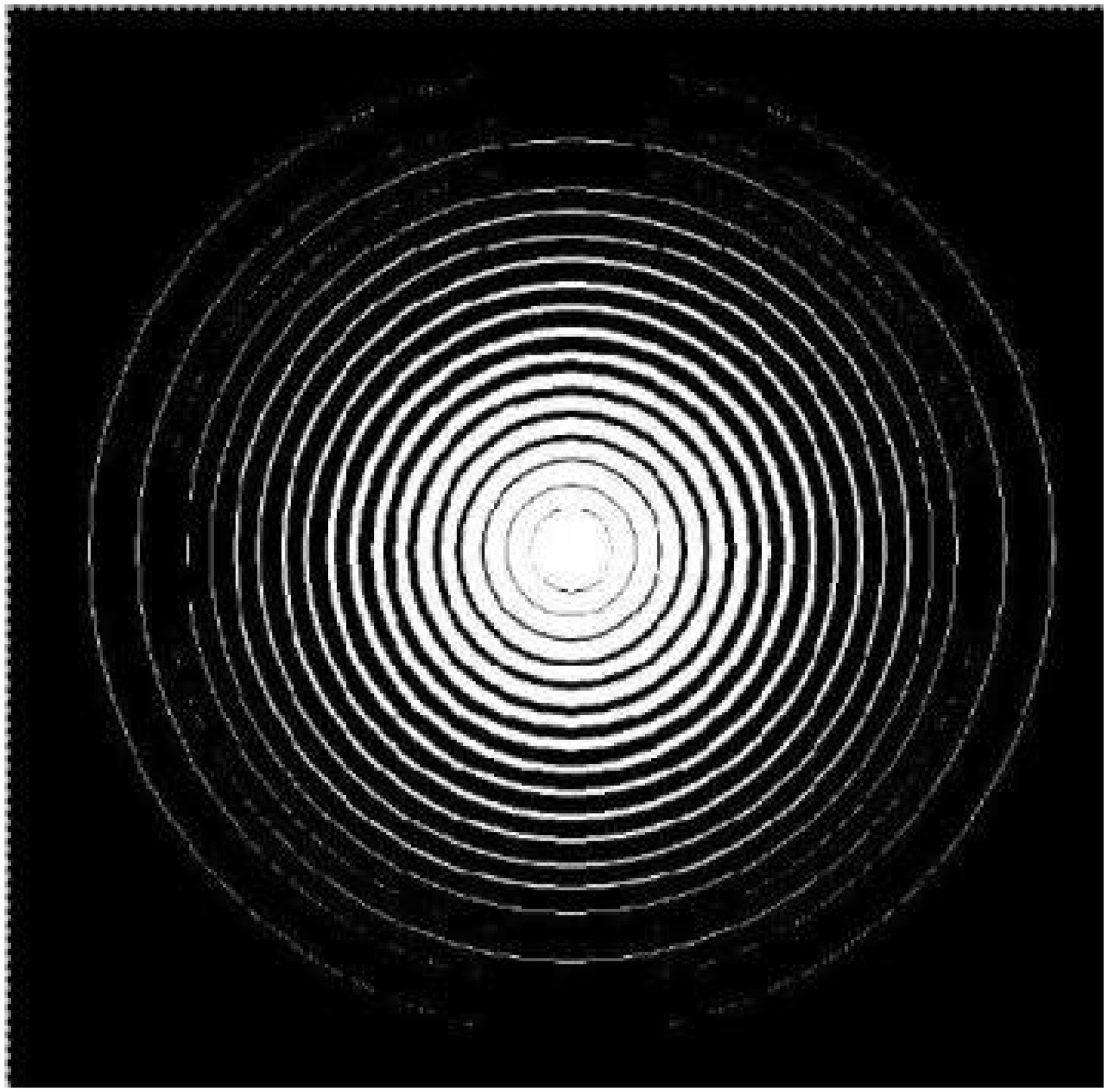} \end{center}
\begin{center} 
\hfill \includegraphics[width=2.0in]{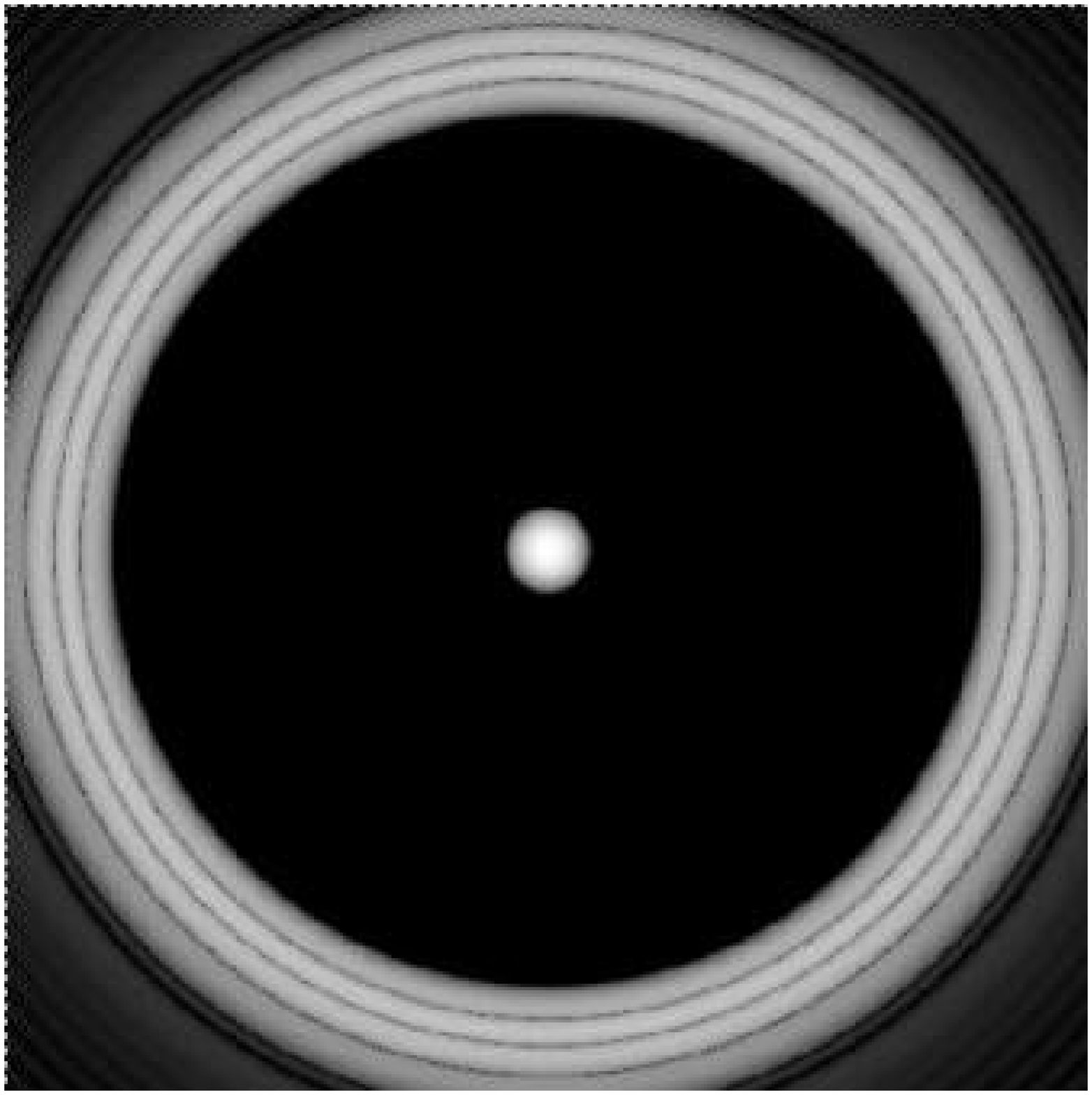}
\hfill \includegraphics[width=2.5in]{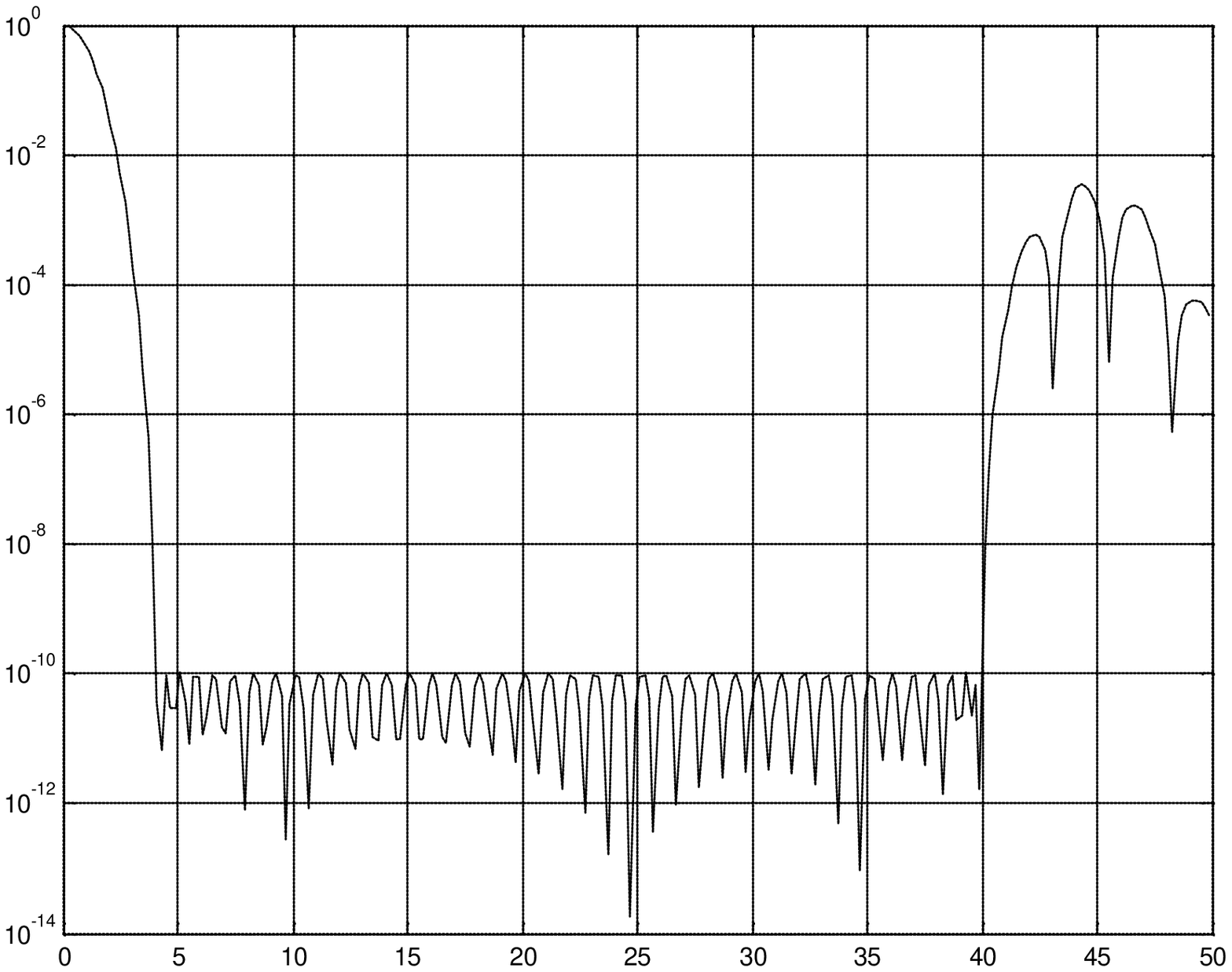}
\hfill ~ 
\end{center}
\caption{{\em Top.} A concentric-ring mask designed to provide high-contrast,
$10^{-10}$, from $\ld = 4$ to $\ld=40$.  Throughput is $0.131 = 16.62\%$.
{\em Bottom.} The associated psf.
}
\label{fig:fig3}
\end{figure}

Figure \ref{fig:fig2} shows the psf for the mask designed 
with $\rhonaught = 4$ and $\rhoone=60$ and three different spider
configurations.  The configuration consisting of 180 spiders each spanning
0.0025 radians completely preserves the full high-contrast region.
Note that this spider fills, in total, $180 \times 0.0025 = 0.45$ radians,
which is $7.2\%$ of the open area.

\begin{figure}
\begin{center} 
\hfill \includegraphics[width=2.0in]{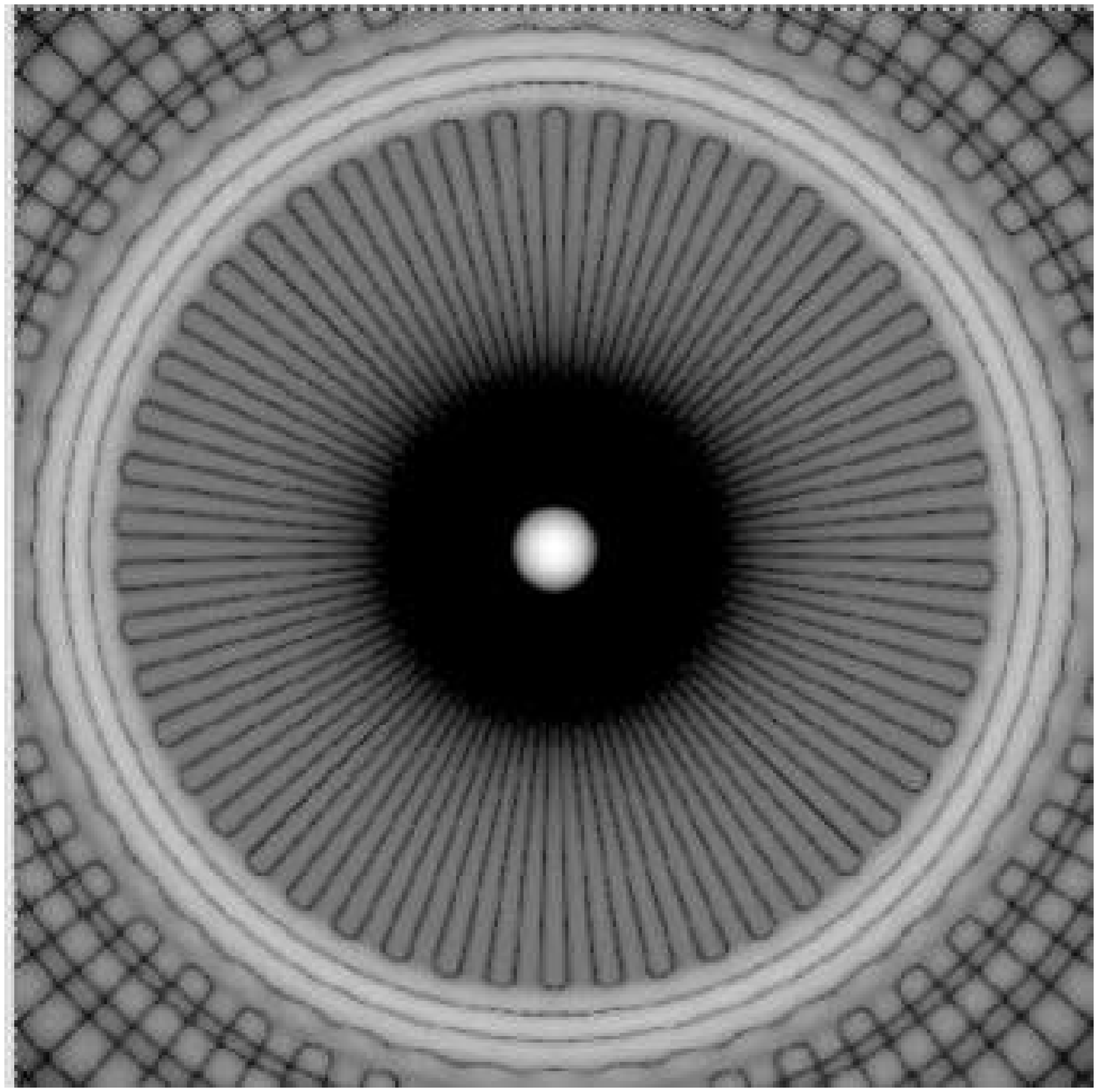}
\hfill \includegraphics[width=2.5in]{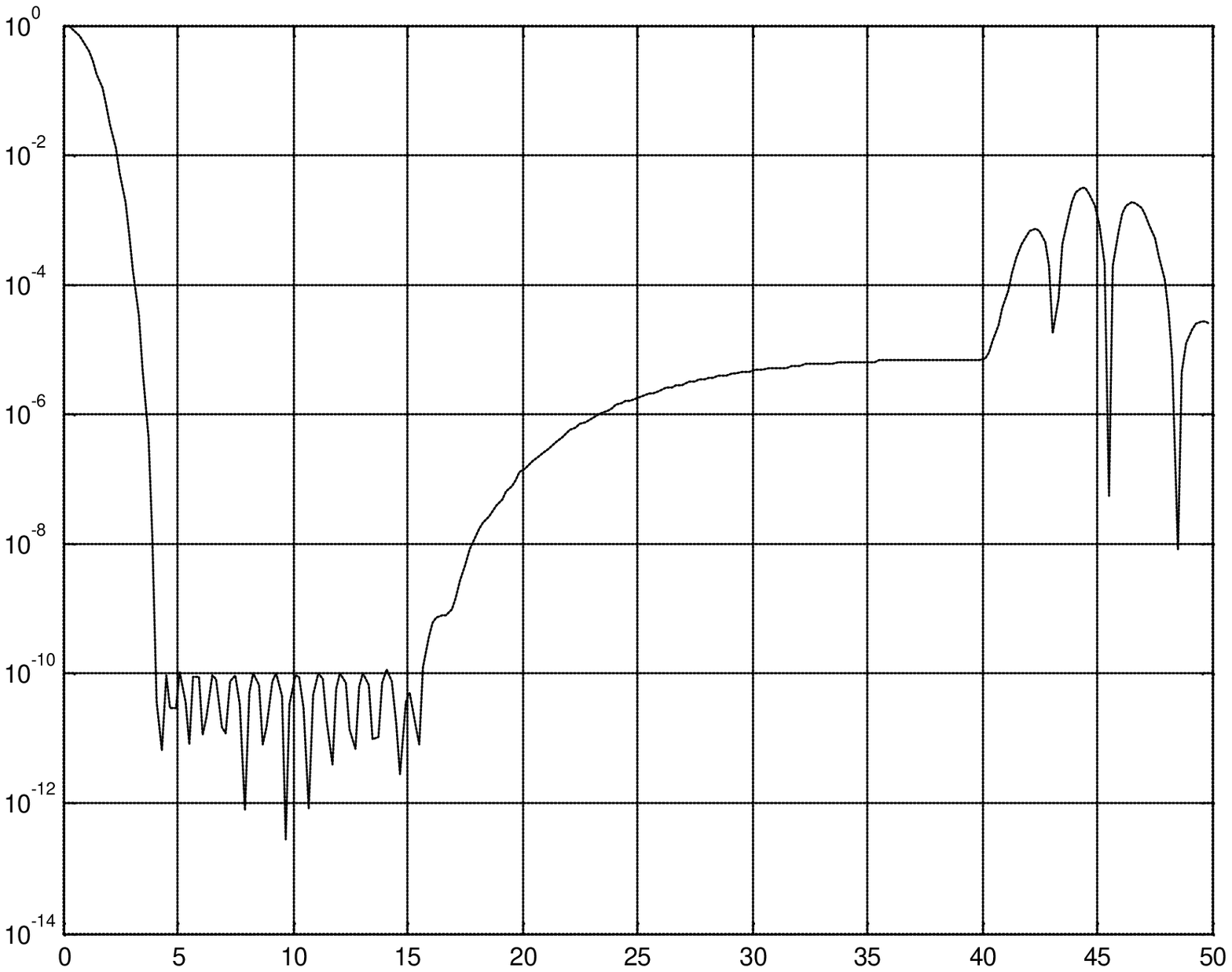}
\hfill ~ \\
\hfill \includegraphics[width=2.0in]{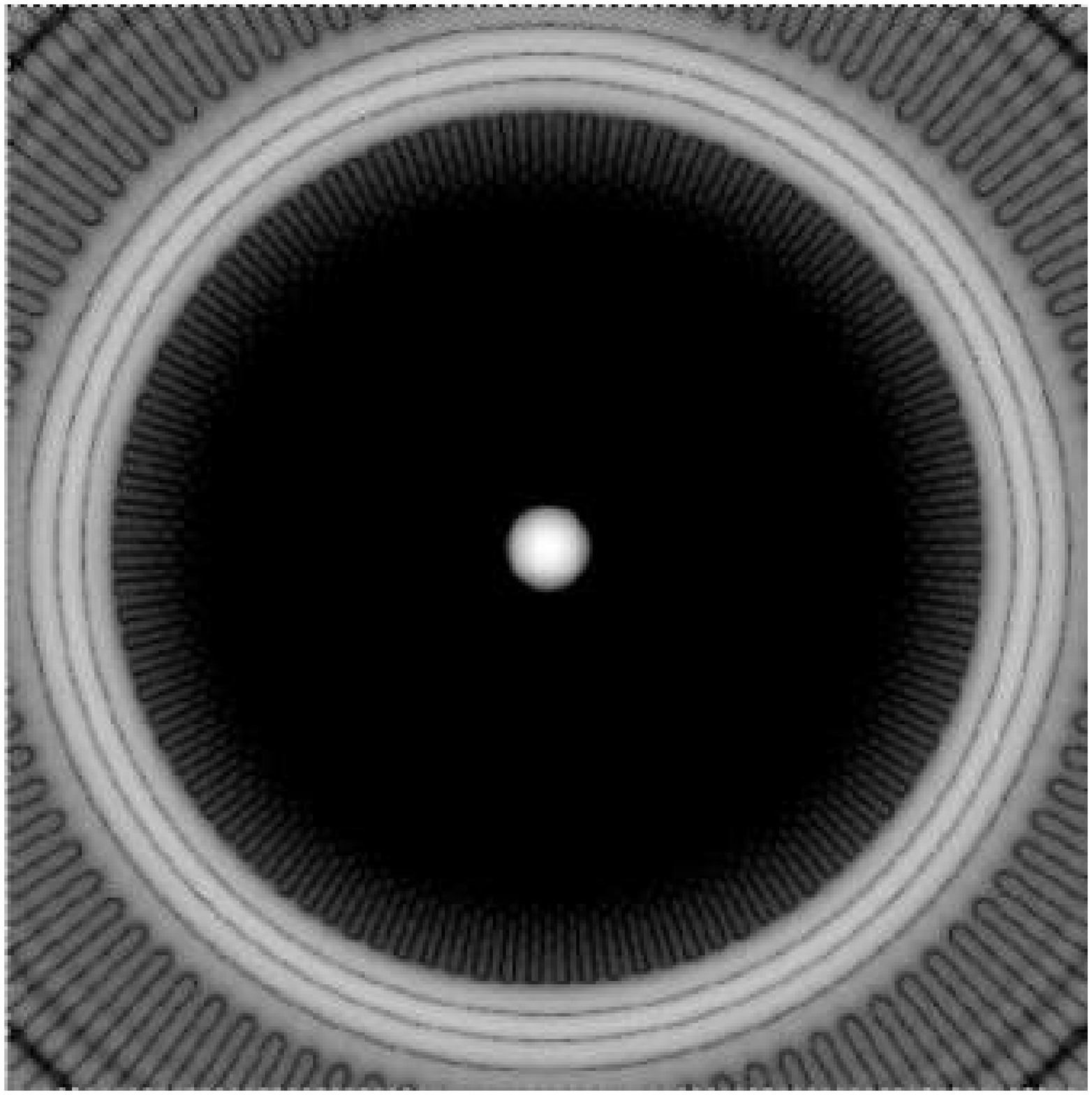}
\hfill \includegraphics[width=2.5in]{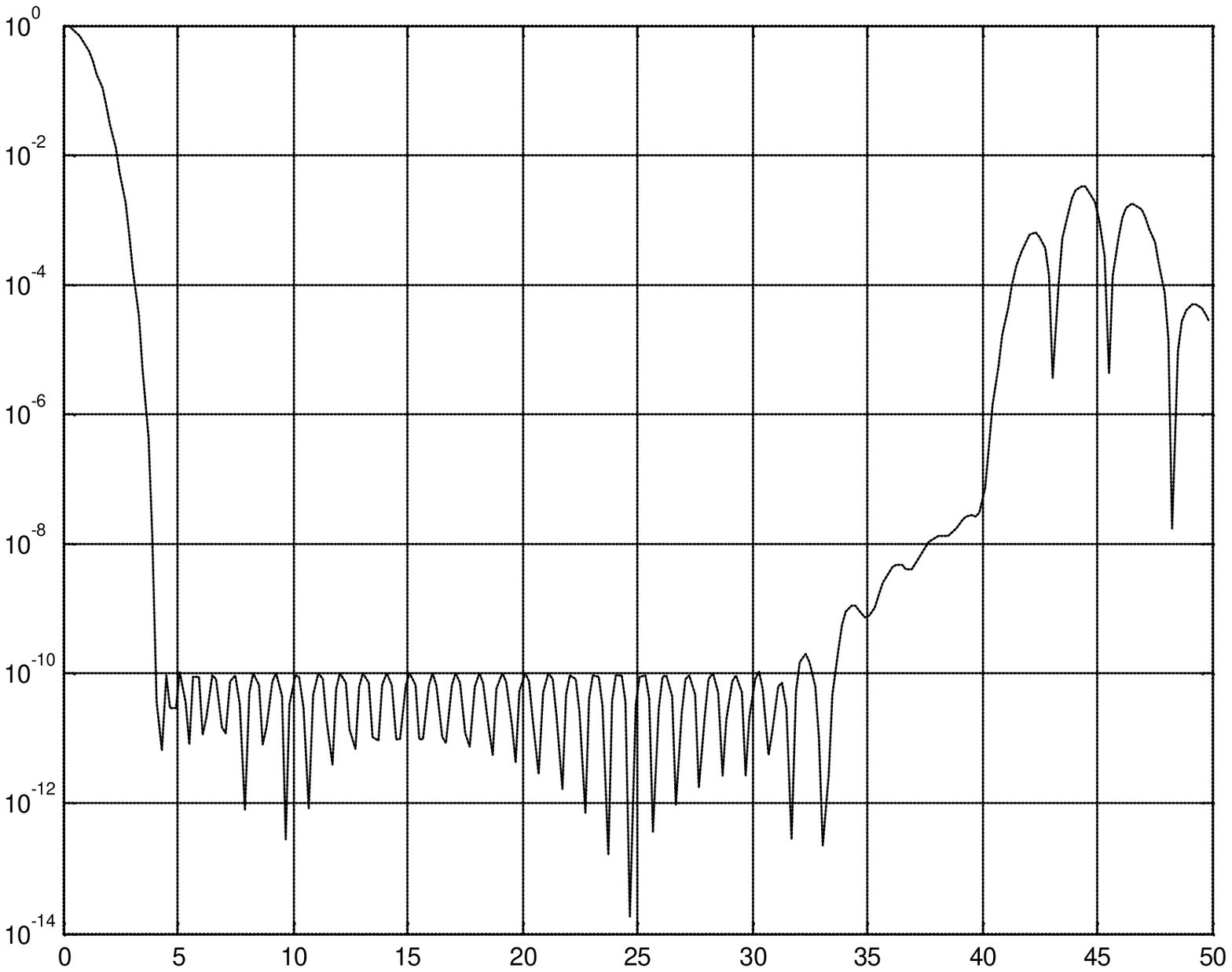}
\hfill ~ \\
\hfill \includegraphics[width=2.0in]{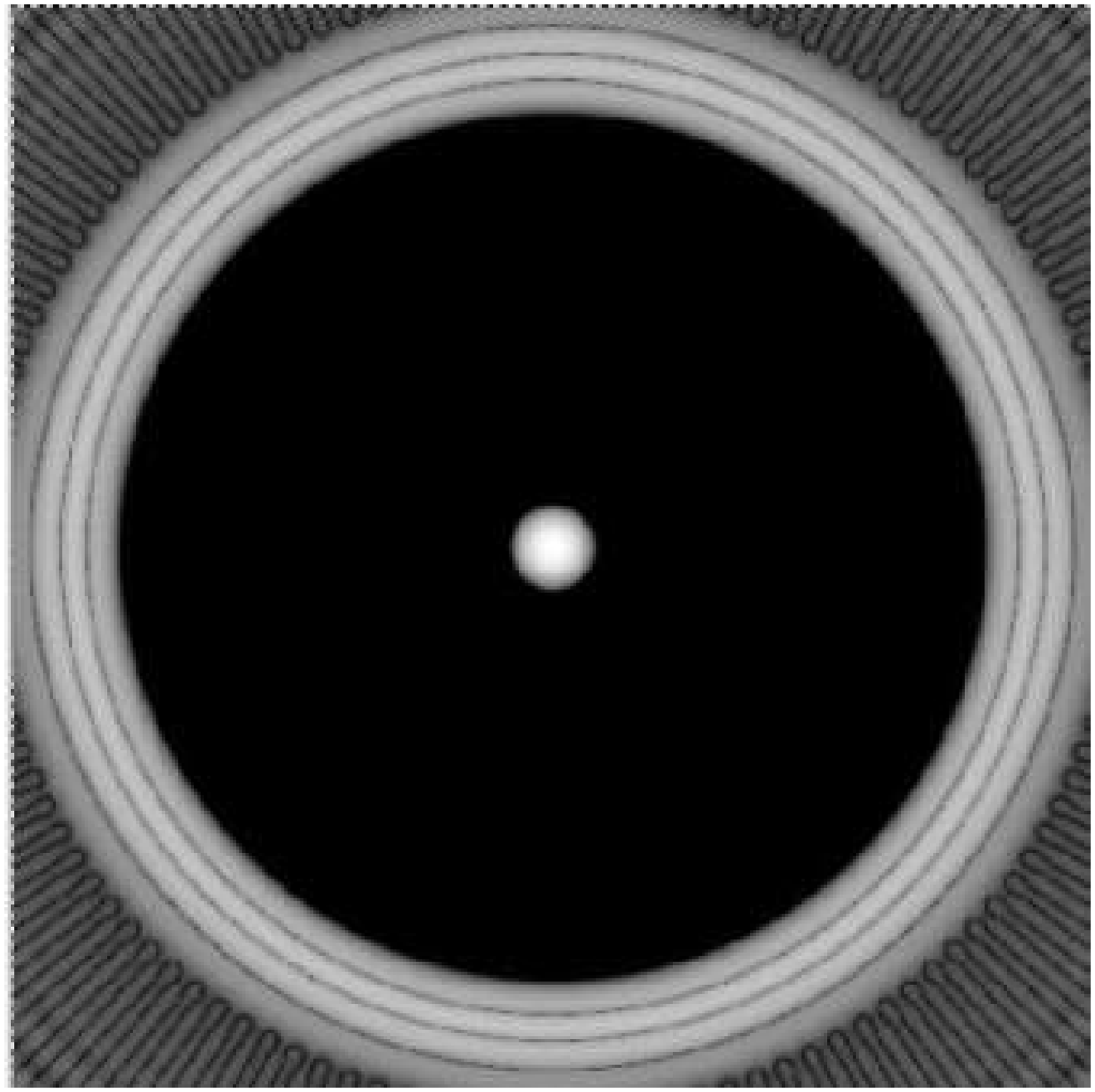}
\hfill \includegraphics[width=2.5in]{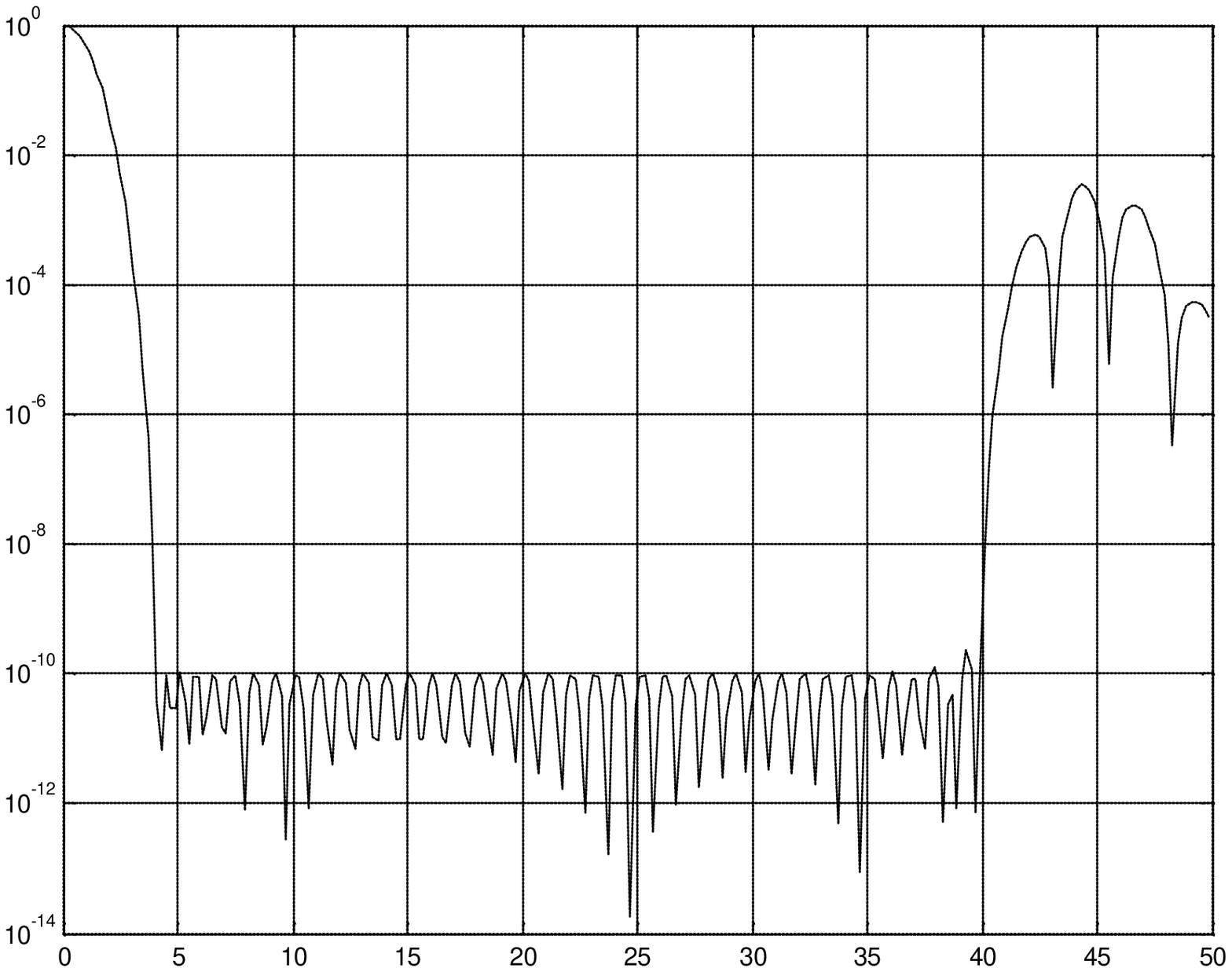}
\hfill ~ 
\end{center}
\caption{Psf for mask in Figure \ref{fig:fig3} with spiders.  
{\em Top Row.} $50$ spiders each spanning $0.01$ radians.
{\em Second Row.} $100$ spiders each spanning $0.005$ radians.
{\em Third Row.} $120$ spiders each spanning $0.003$ radians.
}
\label{fig:fig4}
\end{figure}

Figure \ref{fig:fig4} shows the psf for the mask designed 
with $\rhonaught = 4$ and $\rhoone=40$ and three different spider
configurations.  The configuration consisting of 120 spiders each spanning
0.003 radians completely preserves the full high-contrast region.
Note that this spider fills, in total, $120 \times 0.003 = 0.36$ radians,
which is $5.7\%$ of the open area.

\begin{figure}
\begin{center} \includegraphics[width=2.5in]{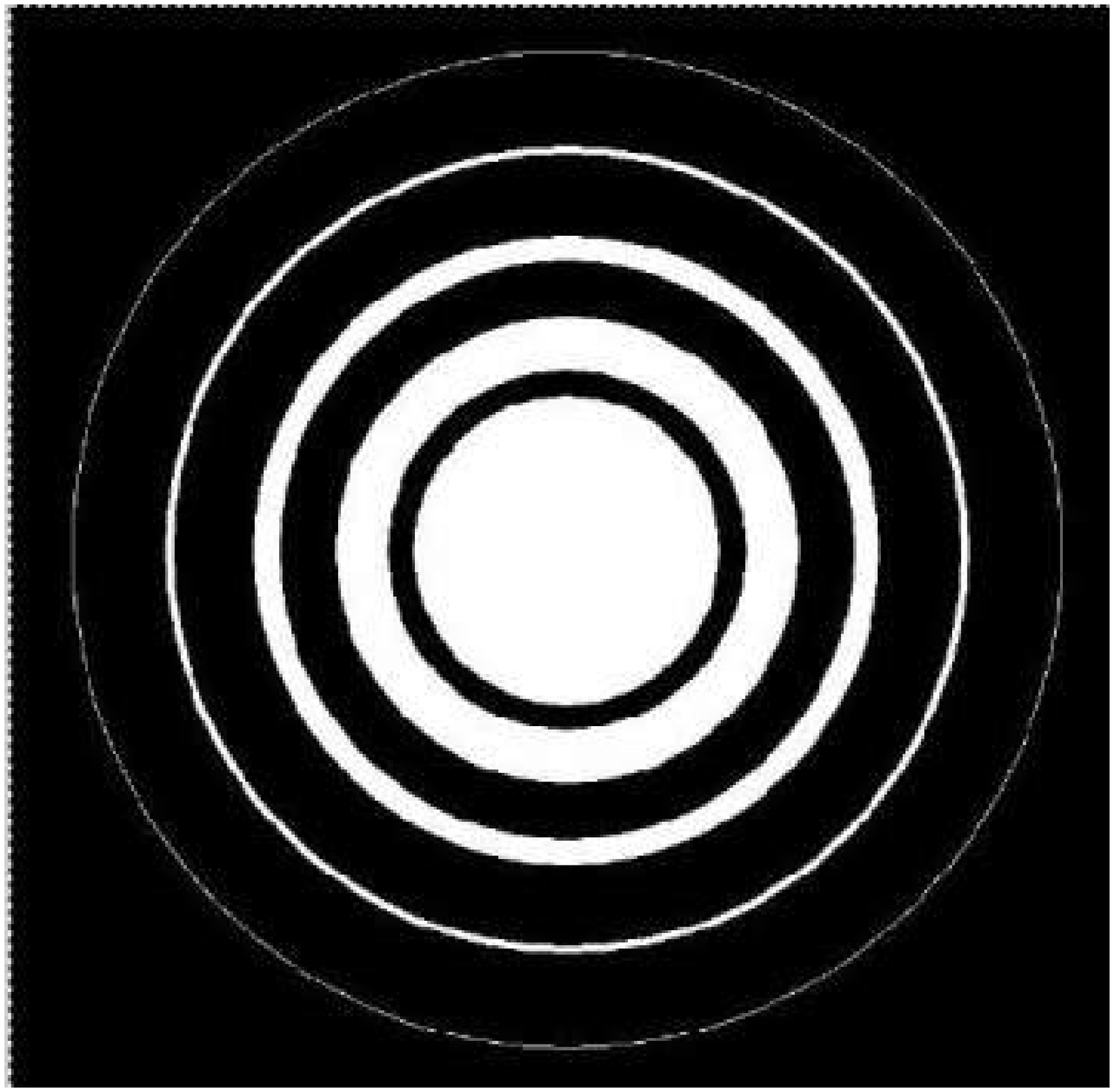} \end{center}
\begin{center} 
\hfill \includegraphics[width=2.0in]{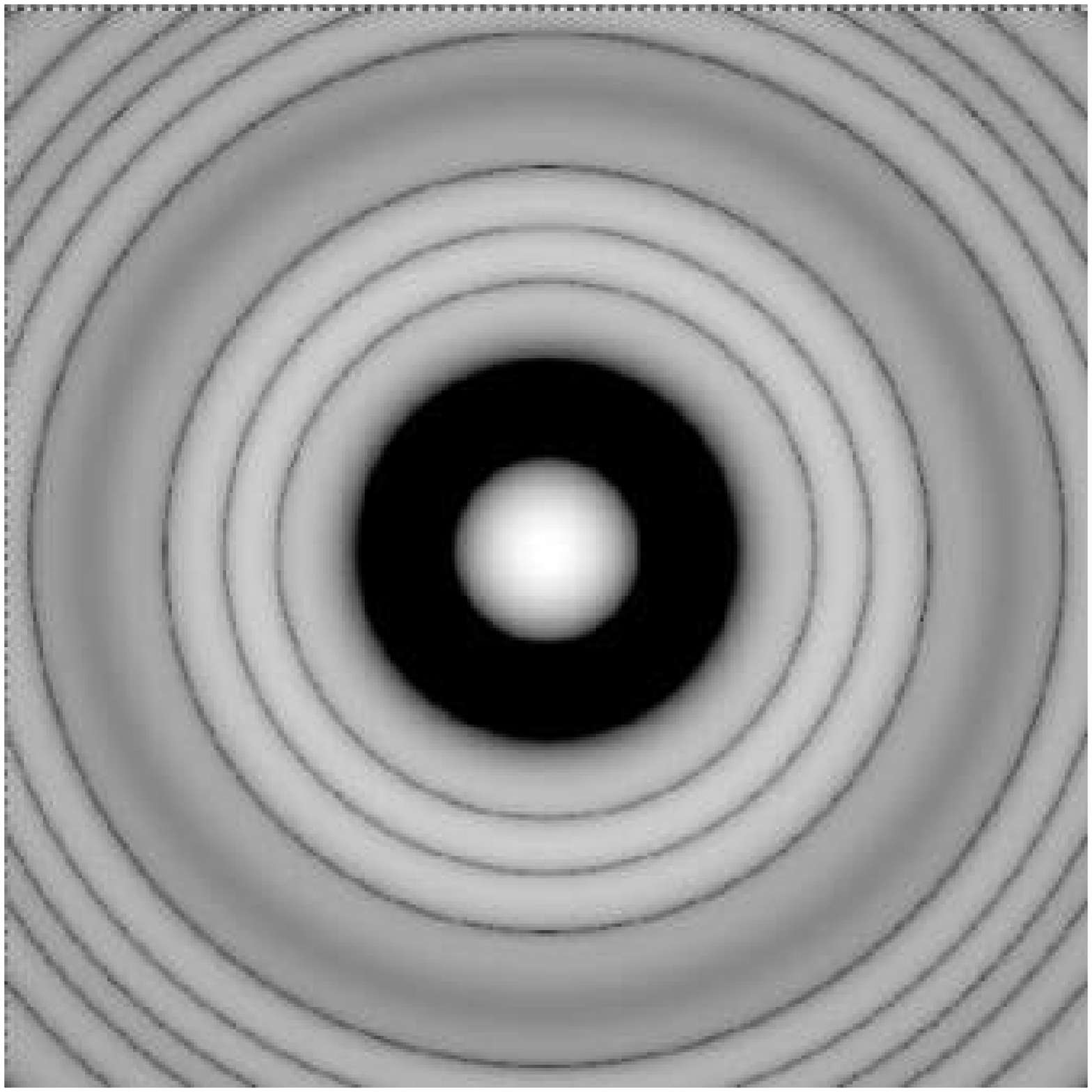}
\hfill \includegraphics[width=2.5in]{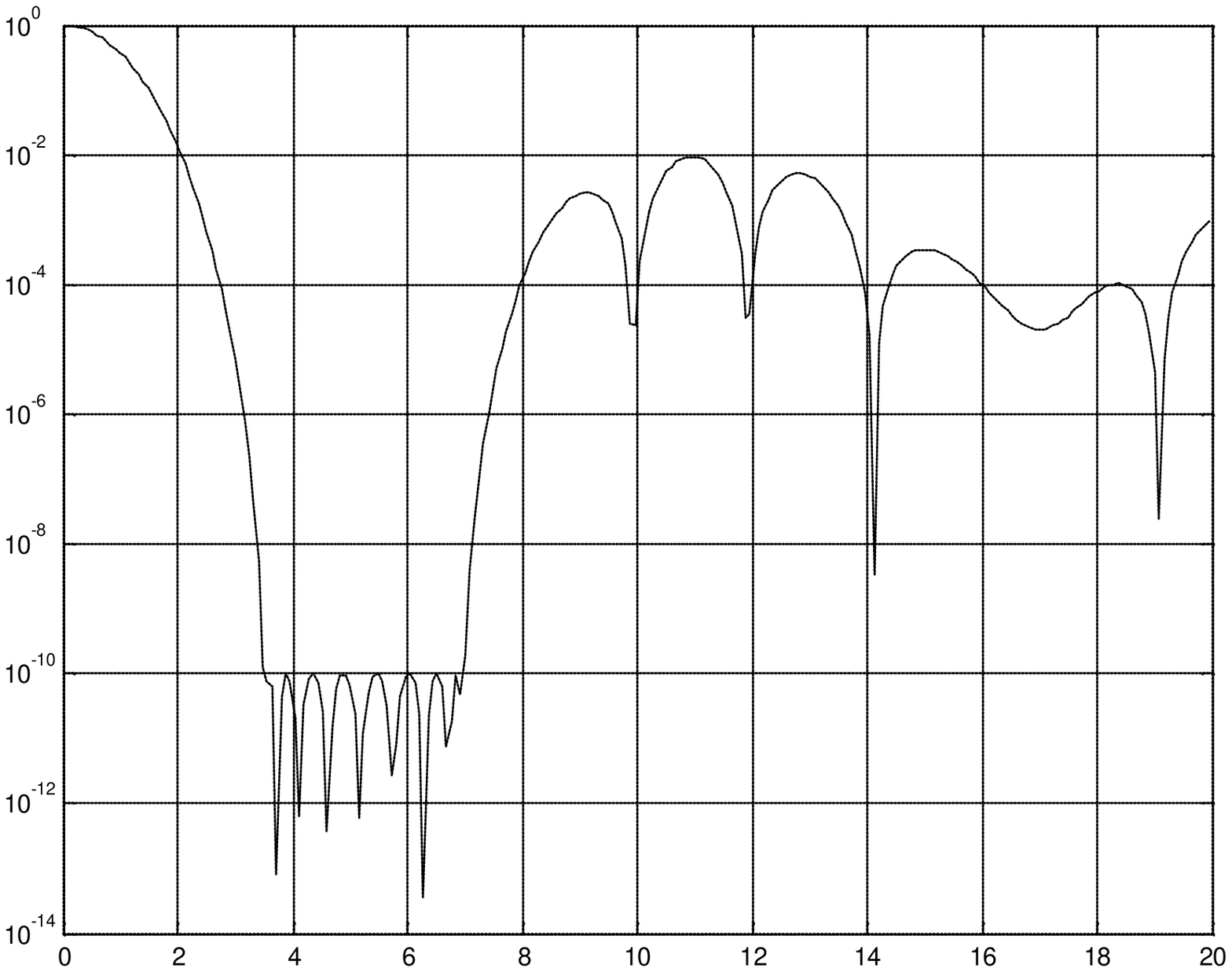}
\hfill ~ 
\end{center}
\caption{{\em Top.} A concentric-ring mask designed to provide high-contrast,
$10^{-10}$, from $\ld = 3.5$ to $\ld=7$.  Throughput is $0.184 = 23.5\%$.
{\em Bottom.} The associated psf.
}
\label{fig:fig5}
\end{figure}

For the mask designed 
with $\rhonaught = 3.5$ and $\rhoone=7$ (shown in Figure \ref{fig:fig5}),
$50$ spider vanes suffice to preserve the high-contrast of the dark zone.
The resulting psf is indistinguishable from the one shown in Figure
\ref{fig:fig5}.  If each spider spans $0.01$ radians
then they fill, in total, $50 \times 0.01 = 0.5$ radians,
which is $8.0\%$ of the open area.

\begin{figure}
\begin{center} \includegraphics[width=2.5in]{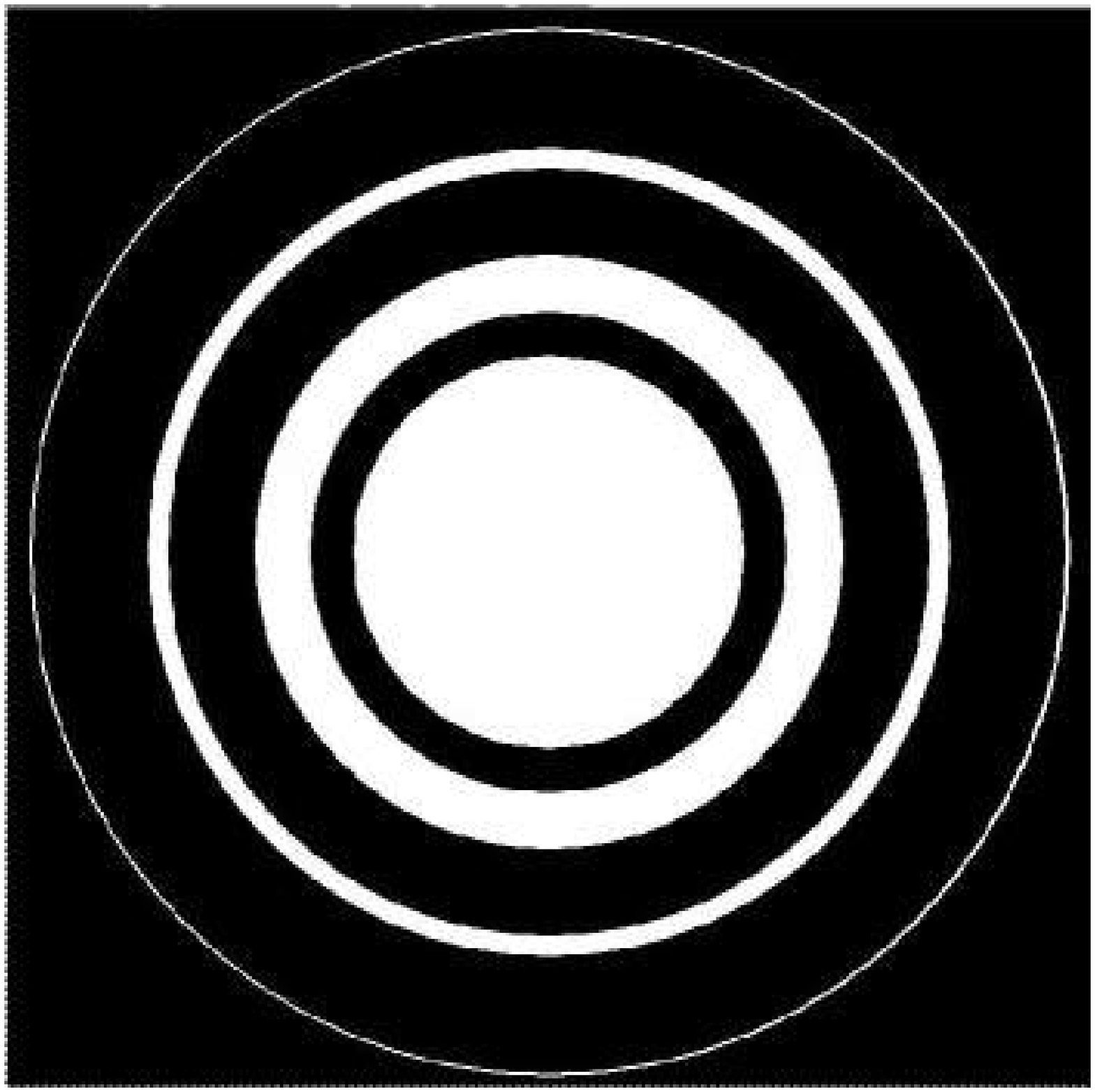} \end{center}
\begin{center} 
\hfill \includegraphics[width=2.0in]{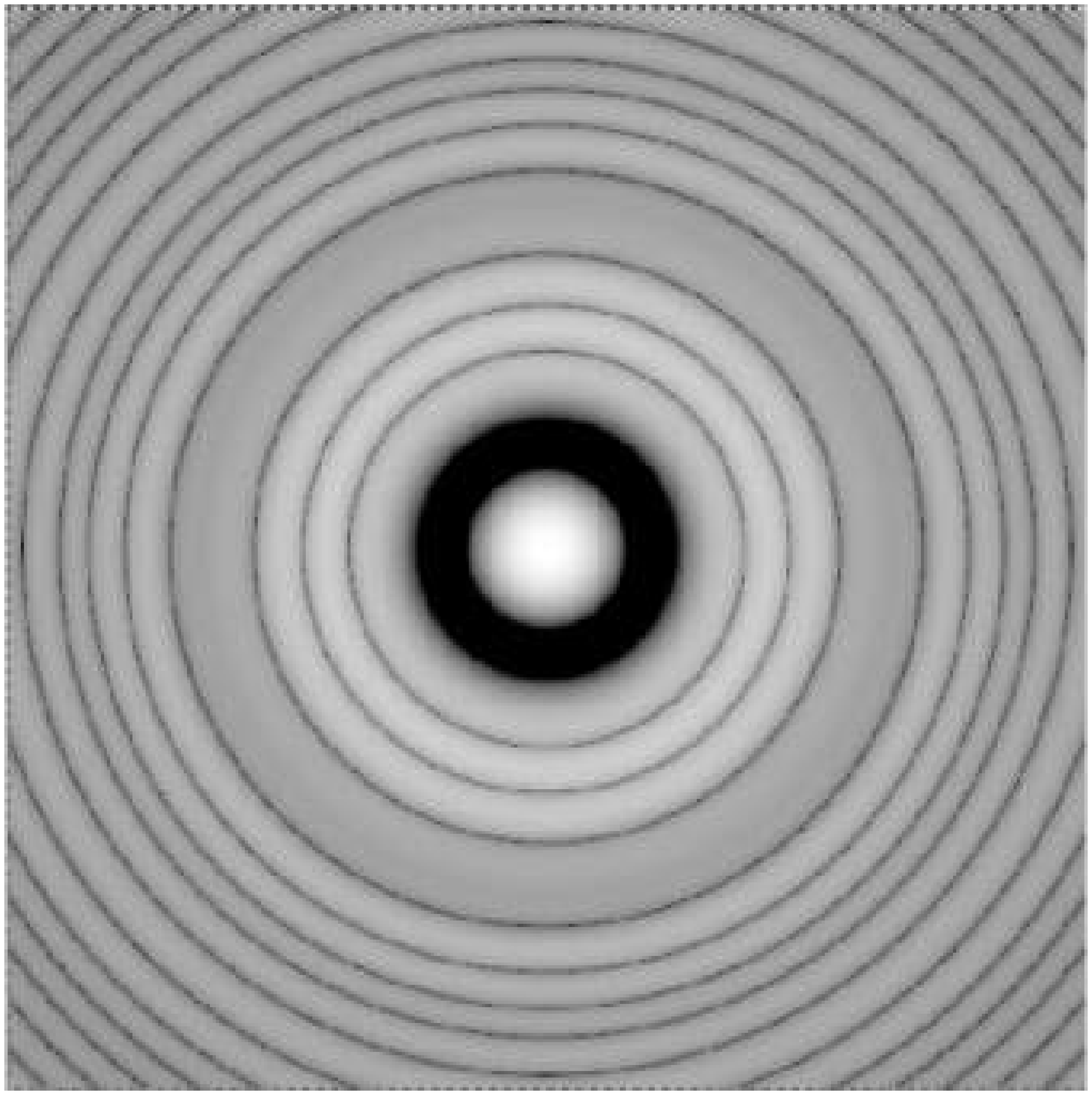}
\hfill \includegraphics[width=2.5in]{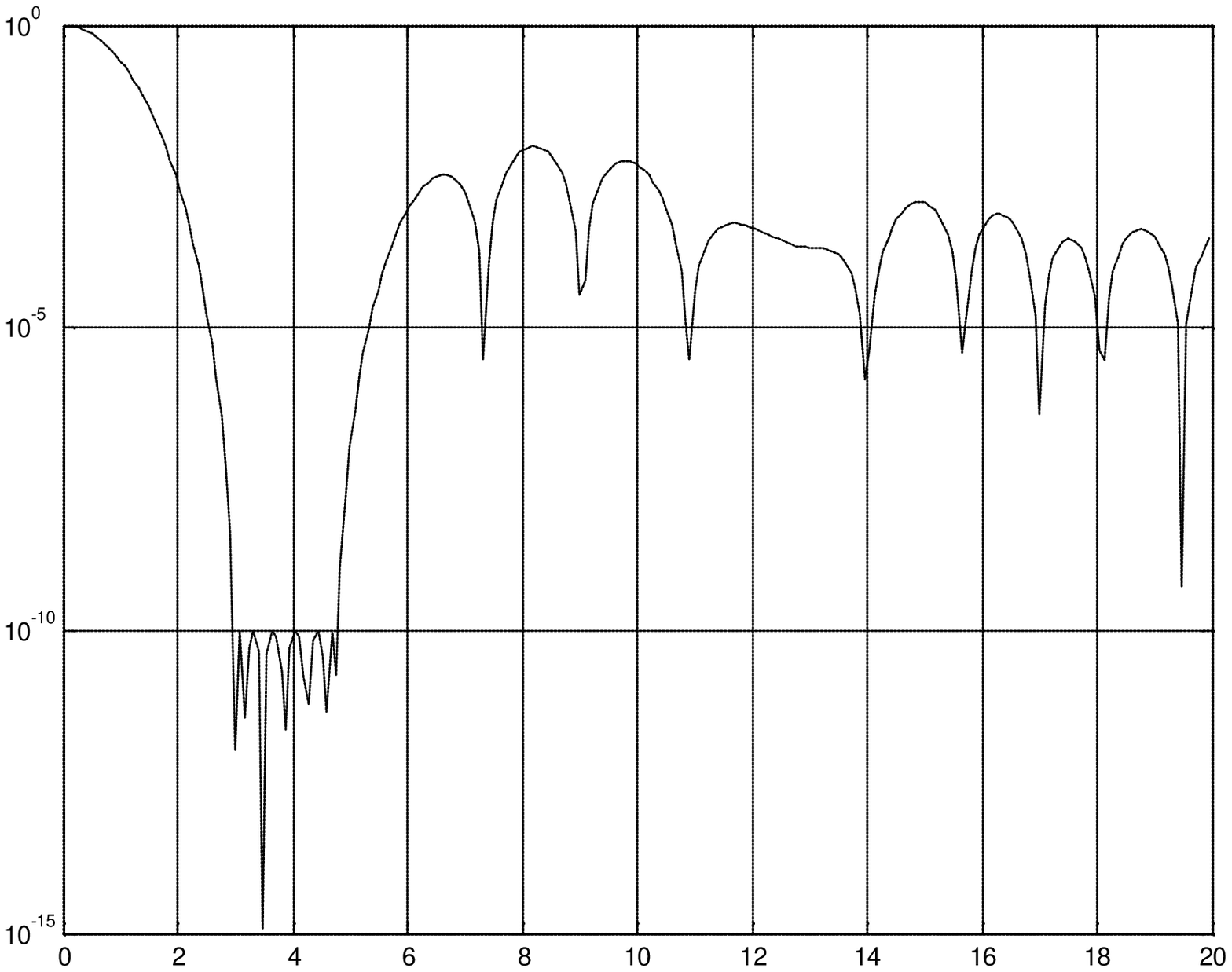}
\hfill ~ 
\end{center}
\caption{{\em Top.} A concentric-ring mask designed to provide high-contrast,
$10^{-10}$, from $\ld = 3.0$ to $\ld=4.8$.  Throughput is $0.251 = 31.9\%$.
{\em Bottom.} The associated psf.
}
\label{fig:fig10}
\end{figure}

For the mask designed 
with $\rhonaught = 3.0$ and $\rhoone=4.8$ (shown in Figure \ref{fig:fig10}),
again
$50$ spider vanes suffice to preserve the high-contrast of the dark zone and
the resulting psf is indistinguishable from the one shown in the earlier
figure.

None of the masks presented have a central obstruction.  Hence, there are
many spiders all coming to a point at the center.  This central area is
probably the hardest part of the mask to manufacture.  One probably needs
to add a small bead at the center as an attachment point for the spiders.
Using the same inner and outer working distances ($3.0$ and $4.8$) as for the
mask in Figure \ref{fig:fig10},
we experimented with adding a constraint to the optimization problem that
forces there to be a small central obstruction.  In the nonlinear optimization
model, the constraint simply amounts to fixing $r_0$ to a specified small
value.  We tried several values and found that we could fix $r_0$ to $0.01$
(i.e., a $2\%$ obstruction) and still find an optimal solution with
essentially an unchanged throughput.

\begin{figure}
\begin{center} \includegraphics[width=2.5in]{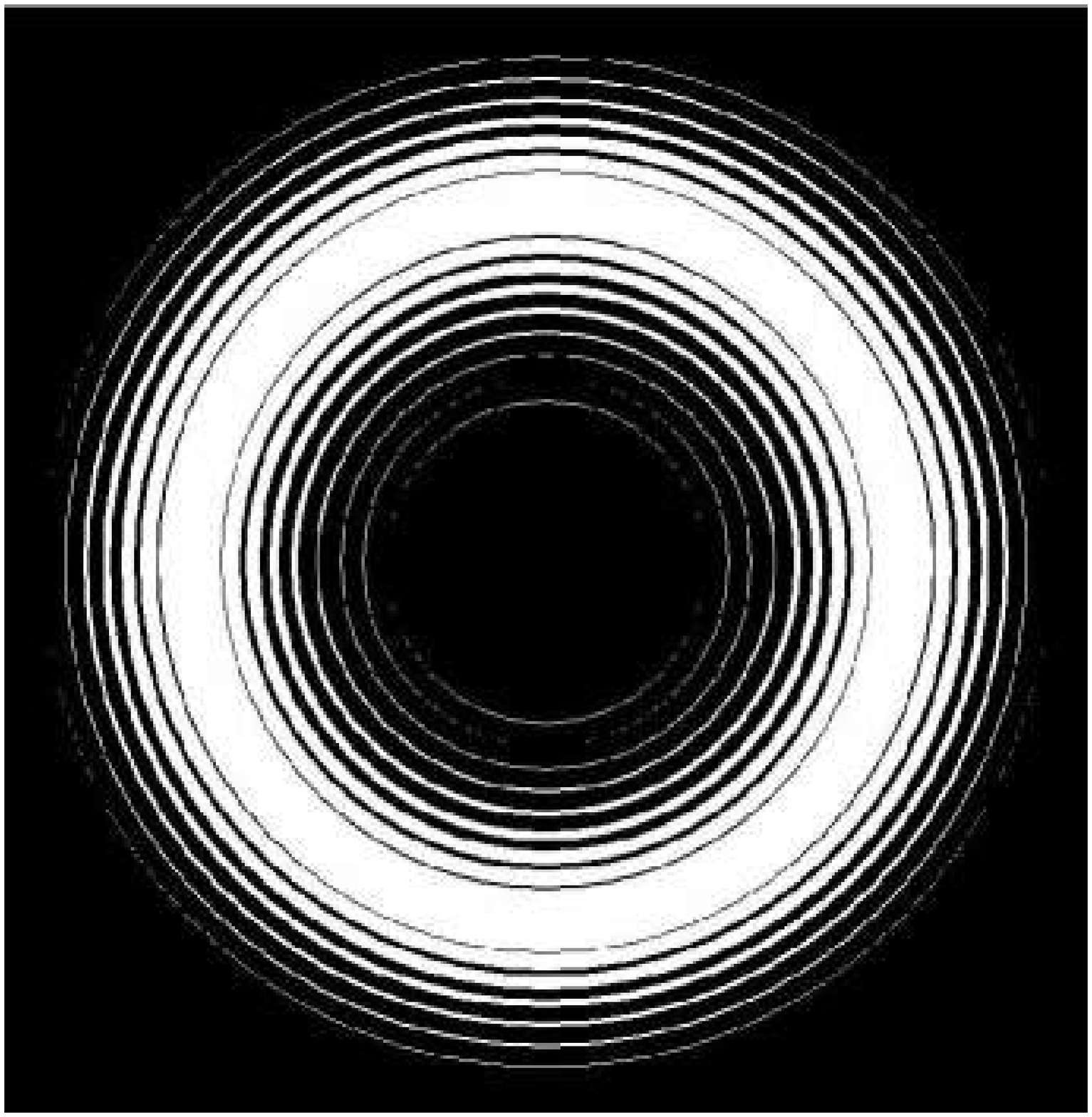} \end{center}
\begin{center} 
\hfill \includegraphics[width=2.0in]{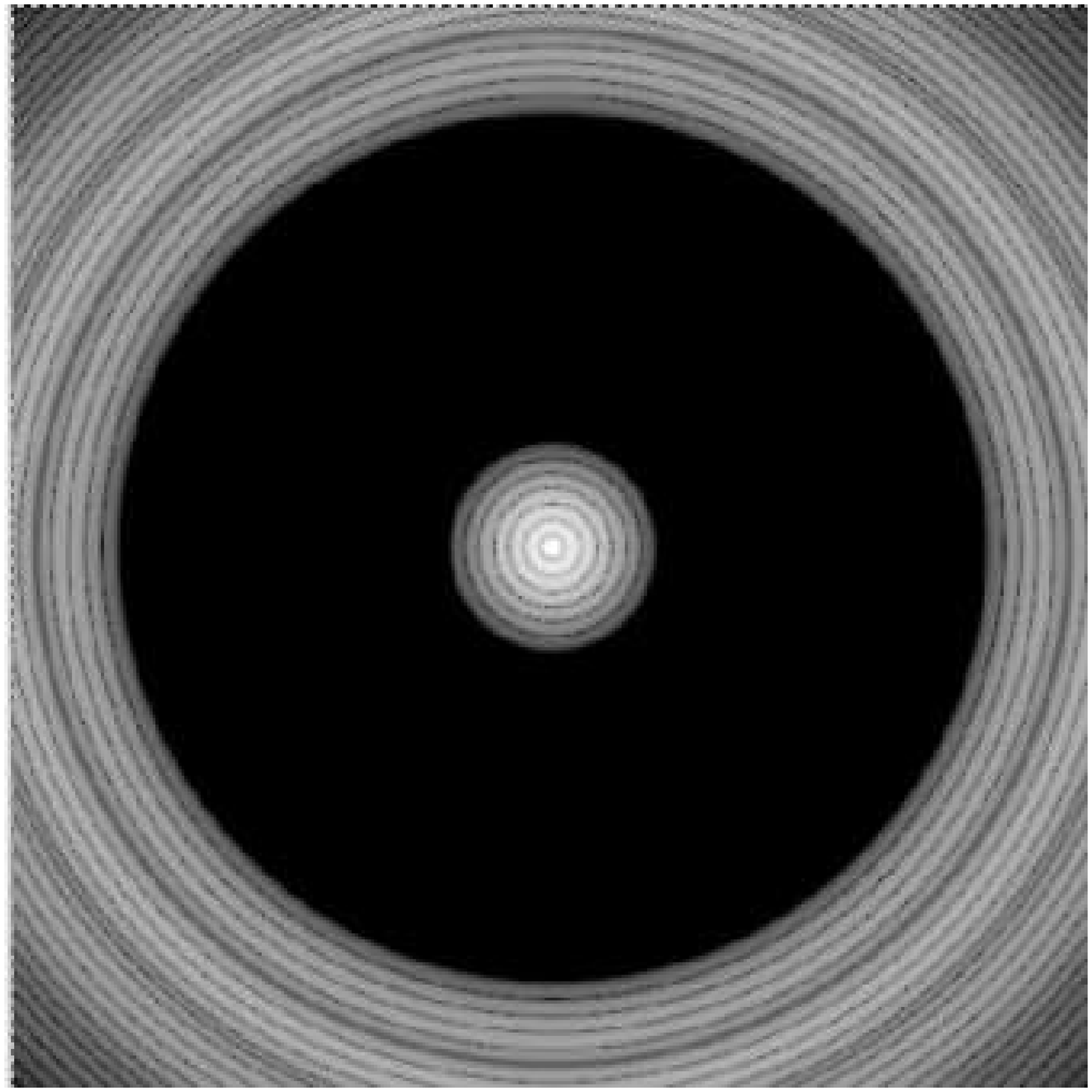}
\hfill \includegraphics[width=2.5in]{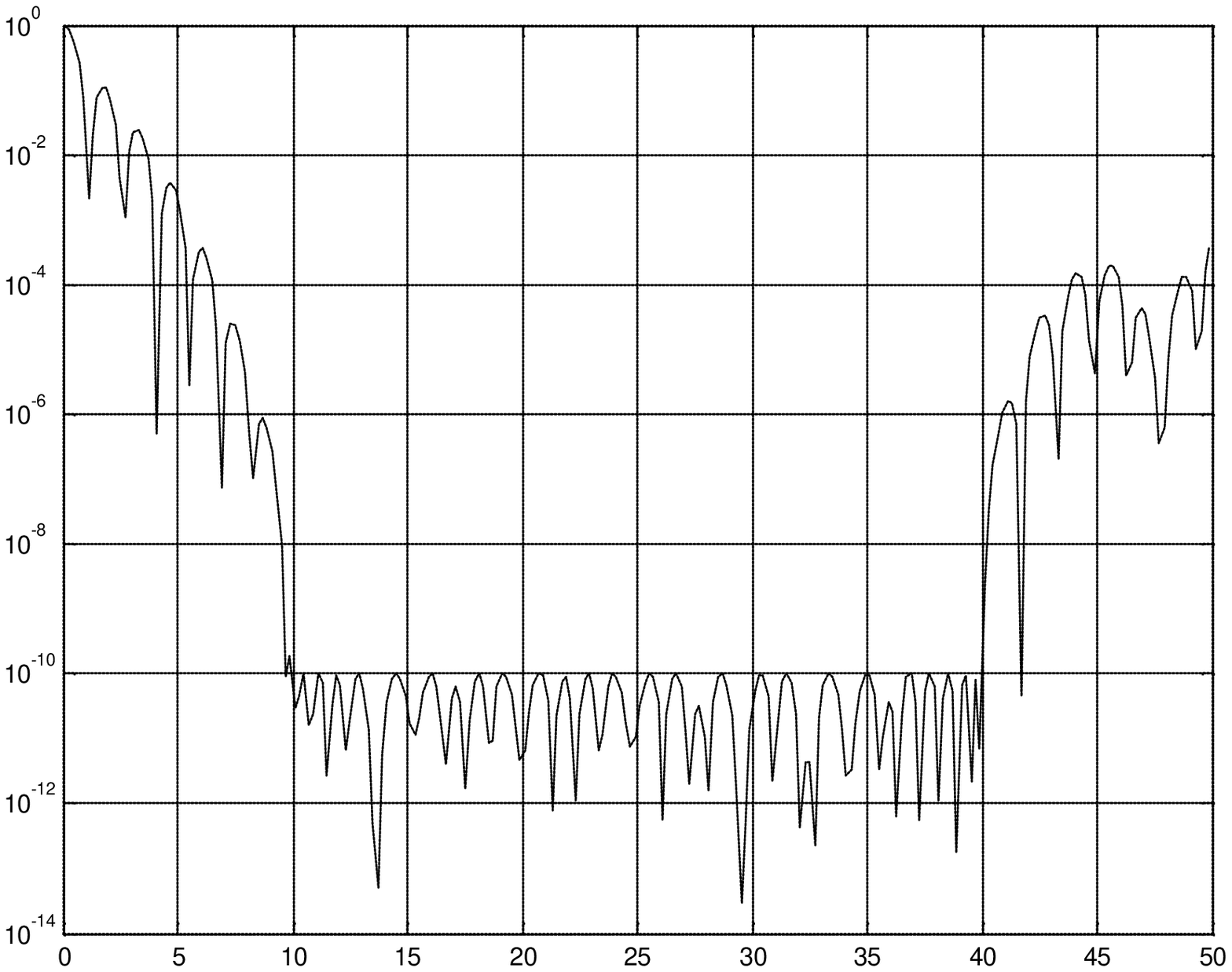}
\hfill ~ 
\end{center}
\caption{{\em Top.} A concentric-ring mask accomodating a large central
obstruction.  High contrast of 
$10^{-10}$, from $\ld = 10.0$ to $\ld=40$.  
{\em Bottom.} The associated psf.
}
\label{fig:fig30}
\end{figure}

\begin{figure}
\begin{center} 
\includegraphics[width=3.0in]{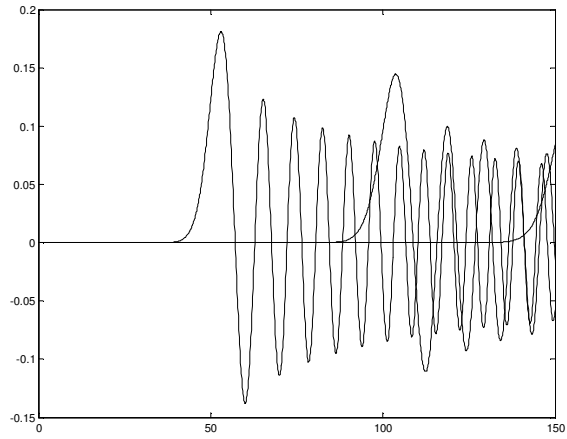}
\end{center}
\caption{The Bessel functions $J_{50}$, $J_{100}$, and $J_{150}$.
They first reach $10^{-5}$ at $35.2$, $81.0$, and $128.1$, respectively.
}
\label{fig:fig8}
\end{figure}

It is a trivial matter to make designs with larger central obstructions
to accomodate, say, a secondary mirror in an on-axis design.  However, once
the central obstruction becomes large, one loses the ability to get tight
inner working distances.  Figure \ref{fig:fig30} shows a design for a $31\%$
central obstruction, an inner working distance of $\rhonaught=10$ 
and an outer working distance of $\rhoone=40$.  
With $\rhonaught=9$, the linear optimization problem has a
solution but not the nonlinear one.  With
$\rhonaught=8$, not even the linear optimization problem has a solution.

Some TPF concepts involve an elliptical pupil geometry since this might
provide a means to achieve improved angular resolution in realizable
rocket fairings.  The designs presented in this paper 
are given in unitless variables.  When re-unitizing, a different
scale can be used for the $x$ and $y$ directions.  In this way, these
designs can be applied directly to elliptical pupils.  Of course, the 
high-contrast region of the psf will also be elliptical with the short
axis of the psf corresponding to the long axis of the pupil.

\section{Final Remarks}

There are two parts to TPF: discovery and characterization.  Discovery refers
to the simple act of looking for exosolar planets.  Characterization refers to
the process of learning as much as possible about specific planets after they
have been discovered.  The masks presented here are intended primarily for
discovery since a single exposure with these masks
can discover a planet in any orientation relative to its star.  
However, once a planet is found and its orientation is known,
some of the asymmetric masks presented in previous papers will be used for
photometry and spectroscopy as they have higher single-exposure throughput.


An alternative to pupil masks is to use a traditional coronagraph, which
consists of an image plane mask followed by a Lyot stop in a reimaged pupil
plane.  Recently, \citet{ref:kuchner} have developed band-limited image-plane
masks that achieve the desired contrast to within $3 \ld$.  However, this 
approach suffers from sensitivity to pointing accuracy.  Nonetheless, the
approach is promising.  In the future, we plan to consider combining pupil
masks with image masks and Lyot stops to make a hybrid design that hopefully
will provide a design achieving the desired contrast with
an even smaller inner working distance.

In this paper we have only considered scalar electric fields.  We leave
the important and more complex issue of how to treat vector 
electric fields, i.e.  polarized light, to future work.

{\bf Acknowledgements.}
We would like to express our gratitude to our colleagues 
on the Ball Aerospace and Technology TPF team.  
We benefited greatly from the many enjoyable and stimulating
discussions.  
This work was partially performed for the Jet Propulsion Laboratory, California
Institute of Technology, 
sponsored by the National Aeronautics and Space Administration as part of
the TPF architecture studies and also under contract number 1240729.
The first author received support from the NSF (CCR-0098040) and
the ONR (N00014-98-1-0036).

\bibliography{../../lib/refs}   

\begin{thebibliography}{9}
\expandafter\ifx\csname natexlab\endcsname\relax\def\natexlab#1{#1}\fi
\expandafter\ifx\csname url\endcsname\relax
  \def\url#1{{\tt #1}}\fi

\bibitem[Arfken and Weber(2000)]{AW00}
G.B. Arfken and H.J. Weber.
\newblock {\em Mathematical Methods for Physicists}.
\newblock Harcourt/Academic Press, 5th edition, 2000.

\bibitem[Brown et~al.(2002)Brown, Burrows, Casertano, Clampin, Eggets, Ford,
  Jucks, Kasdin, Kilston, Kuchner, Seager, Sozzetti, Spergel, Traub, Trauger,
  and Turner]{ref:Brown}
R.~A. Brown, C.~J. Burrows, S.~Casertano, M.~Clampin, D.~Eggets, E.B. Ford,
  K.W. Jucks, N.~J. Kasdin, S.~Kilston, M.~J. Kuchner, S.~Seager, A.~Sozzetti,
  D.~N. Spergel, W.~A. Traub, J.~T. Trauger, and E.~L. Turner.
\newblock The 4-meter space telescope for investigating extrasolar earth-like
  planets in starlight: {T}{P}{F} is {H}{S}{T}2.
\newblock In {\em Proceedings of SPIE: Astronomical Telescopes and
  Instrumentation}, number~14 in 4860, 2002.

\bibitem[Kasdin et~al.(2002{\natexlab{a}})Kasdin, Spergel, and
  Littman]{ref:kasdin}
N.~J. Kasdin, D.~N. Spergel, and M.~G. Littman.
\newblock An optimal shaped pupil coronagraph for high contrast imaging, planet
  finding, and spectroscopy.
\newblock {\em submitted to Applied Optics}, 2002{\natexlab{a}}.

\bibitem[Kasdin et~al.(2002{\natexlab{b}})Kasdin, Vanderbei, Spergel, and
  Littman]{KVSL02b}
N.J. Kasdin, R.J. Vanderbei, D.N. Spergel, and M.G. Littman.
\newblock 
  {Optimal Shaped Pupil Coronagraphs for Extrasolar Planet Finding}.
\newblock In {\em Proceedings of SPIE Conference on Astronomical Telescopes and
  Instrumentation}, number~44 in 4860, 2002{\natexlab{b}}.

\bibitem[Kasdin et~al.(2003)Kasdin, Vanderbei, Spergel, and Littman]{KVSL02}
N.J. Kasdin, R.J. Vanderbei, D.N. Spergel, and M.G. Littman.
\newblock 
  {Extrasolar Planet Finding via Optimal Apodized and Shaped Pupil
  Coronagraphs}.
\newblock {\em Astrophysical Journal}, 2003.
\newblock To appear.

\bibitem[Kuchner and Traub(2002)]{ref:kuchner}
M.~J. Kuchner and W.~A. Traub.
\newblock A coronagraph with a band-limited mask for finding terrestrial
  planets.
\newblock {\em The Astrophysical Journal}, \penalty0 (570):\penalty0 900+,
  2002.

\bibitem[Spergel(2000)]{ref:Spergel}
D.~N. Spergel.
\newblock A new pupil for detecting extrasolar planets.
\newblock {\em astro-ph/0101142}, 2000.

\bibitem[Vanderbei(1999)]{SOR9708}
R.J. Vanderbei.
\newblock {L}{O}{Q}{O} user's manual---version 3.10.
\newblock {\em Optimization Methods and Software}, 12:\penalty0 485--514, 1999.

\bibitem[Vanderbei(2001)]{Van01}
R.J. Vanderbei.
\newblock {\em Linear Programming: Foundations and Extensions}.
\newblock Kluwer Academic Publishers, 2nd edition, 2001.

\end{thebibliography}
\bibliographystyle{plainnat}   

\clearpage

\clearpage

\end{document}